\documentstyle[emulateapj,10pt,apjfonts,psfig]{article}
\def\snr{G320.4--1.2}
\def\msh{MSH~15--5{\em 2}}
\def\rcw{RCW~89}
\def\psr{B1509--58}
\def\cxo{{\em Chandra}}

\newcommand\HI{H\,{\sc i}}

\def\etal{et~al. }

\lefthead{GAENSLER ET AL}
\righthead{THE X-RAY NEBULA POWERED BY PSR~\psr}

\begin{document}
\title{Chandra Imaging of The X-ray Nebula Powered by Pulsar~\psr}
\submitted{To appear in The Astrophysical Journal, vol 569, 2002 April 20}
\author{B. M. Gaensler,\altaffilmark{1,2,3} 
J. Arons,\altaffilmark{4} 
V. M. Kaspi,\altaffilmark{5,1,6} 
M. J. Pivovaroff,\altaffilmark{7,1}
 N. Kawai\altaffilmark{8,9} and K.~Tamura\altaffilmark{10}}
\altaffiltext{1}{Center for Space Research, Massachusetts Institute of
Technology, 70 Vassar Street, Cambridge, MA 02139}
\altaffiltext{2}{Harvard-Smithsonian
Center for Astrophysics, 60 Garden Street MS-6, Cambridge, MA 02138;
bgaensler@cfa.harvard.edu}
\altaffiltext{3}{Hubble Fellow, Clay Fellow}
\altaffiltext{4}{Department of Astronomy, University of
California, Berkeley, CA 94720}
\altaffiltext{5}{Physics Department, McGill University,
3600 University Street, Montreal, Quebec, Canada}
\altaffiltext{6}{Canada Research Council Chair, Alfred P. Sloan Fellow}
\altaffiltext{7}{Space Sciences Laboratory,
University of California, Berkeley, CA 94720}
\altaffiltext{8}{Department of Physics, Tokyo Institute of Technology, 2-12-1
Ookayama, Meguro-ku, Tokyo 152-8551, Japan}
\altaffiltext{9}{Cosmic Radiation Laboratory, RIKEN,
2-1 Hirosawa, Wako, Saitama 351-0198, Japan}
\altaffiltext{10}{Department of Physics, School of Science,
Nagoya University, Furo-cho,
Chikusa-ku, Nagoya 464-8602, Japan}
\hspace{-5cm}

\begin{abstract}

We present observations with the {\em Chandra X-ray Observatory}\ of
the pulsar wind nebula (PWN) powered by the energetic young
pulsar~\psr. These data confirm the complicated morphology of the
system indicated by previous observations, and in addition reveal
several new components to the nebula. The overall PWN shows a clear
symmetry axis oriented at a position angle
$150^\circ\pm5^\circ$ (north through east),
which we argue corresponds to the pulsar spin axis. We
show that a previously identified radio feature matches well with the
overall extent of the X-ray PWN, and propose the former as the
long-sought radio nebula powered by the pulsar. We further identify a
bright collimated feature, at least $4'$ long, lying along the
nebula's main symmetry axis; we interpret this feature as a physical
outflow from the pulsar, and infer a velocity for this jet $>0.2c$. The
lack of any observed counter-jet implies that the pulsar spin axis is
inclined at $\sim30^\circ$ to the line-of-sight, contrary to previous
estimates made from lower-resolution data.  We also identify
a variety of compact features close to the
pulsar.  A pair of semi-circular X-ray arcs lie $17''$ and $30''$ to
the north of the pulsar; the latter arc shows a highly-polarized radio
counterpart. We show that these features can be interpreted as
ion-compression wisps in a particle-dominated equatorial flow, and use
their properties to infer a ratio of electromagnetic to particle energy
in pairs at the wind shock $\sigma \sim 0.005$, similar to that
seen in the Crab Nebula. We further identify several compact knots seen
very close to the pulsar; we use these to infer $\sigma < 0.003$ at
a separation from the pulsar of 0.1~pc.

\end{abstract}

\keywords{
ISM: individual: (\snr) ---
ISM: jets and outflows ---
pulsars: individual (\psr) ---
stars: neutron ---
supernova remnants ---
X-rays: ISM}

\section{Introduction}
\label{sec_intro}

Pulsars and supernova remnants (SNRs) are both believed to be formed
in supernova explosions. However, there are still fewer than 20 cases
in which convincing associations between a pulsar and a SNR have been
established. These few systems provide important information on the
properties of pulsars and SNRs, and on the relationship and interaction
between them. In particular, in cases
where a pulsar is still inside its associated SNR, the high-pressure
environment can confine and shock the pulsar's relativistic wind. This
generates synchrotron emission, producing an observable pulsar wind nebula
(PWN) which can be used to trace the energy flow away from the pulsar.

For almost 15 years, the Crab Nebula and the Vela SNR were the only
SNRs known to be associated with pulsars.  A third association only
emerged when a 150-ms X-ray, radio and $\gamma$-ray pulsar,
PSR~\psr, was discovered within the SNR~\snr\ (\msh) (\cite{sh82};
\cite{mtd82}; \cite{umw+93}).
The pulsar's spin-parameters can be used to infer a characteristic
age $\tau_c = 1700$~yr, a  spin-down luminosity $\dot{E} = 1.8 \times
10^{37}$~erg~s$^{-1}$ and a surface magnetic field $B_p=1.5\times10^{13}$~G
(\cite{kms+94}), making it one of the youngest, most energetic and
highest-field pulsars known.

\begin{figure*}[hbt]
\vspace{1cm}
\centerline{\psfig{file=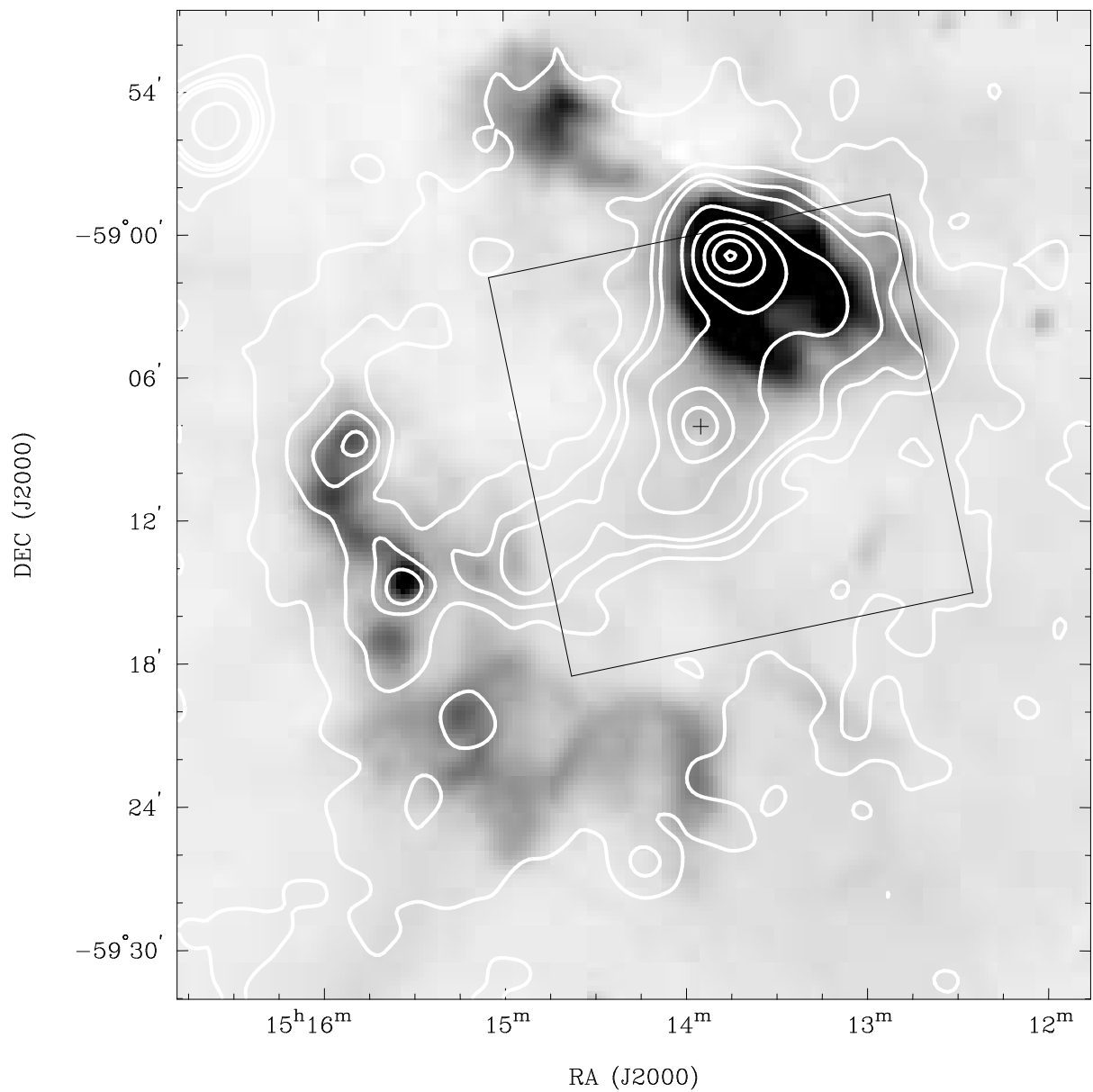,height=16cm,angle=270}}
\caption{A radio/X-ray comparison of \snr. The greyscale corresponds to
843-MHz MOST observations (\protect\cite{wg96}), while the white
contours represent smoothed {\em ROSAT}\ PSPC data 
(\protect\cite{tmc+96}). Contour levels (in arbitrary units) are
at levels of 0.5, 1, 1.5, 2, 5, 10, 20, 30 and 40. The position of
PSR~\protect\psr\ (as determined by G99\protect\nocite{gbm+98})
is marked by a ``$+$'' symbol. The
central box delineates the \protect\cxo\ ACIS-I field-of-view.}
\label{fig_g320_most_rosat}
\end{figure*}

SNR~\snr\ has an unusual radio appearance, consisting of two
well-separated radio regions 
(Fig.~\ref{fig_g320_most_rosat}; \cite{gbm+98}, hereafter G99).  
While the more southerly of these
regions approximates a partial shell, the northern component is bright,
centrally concentrated, and has an unusual ring of radio clumps near
its center. While it had been previously suggested that \snr\ consisted of
two or three separate SNRs, G99
have used \HI\ absorption measurements to confirm that this source is
indeed a single SNR at a distance of $5.2\pm1.4$~kpc. 
Assuming standard parameters for the interstellar medium (ISM)
and for the supernova explosion, an age of $\sim$6--20~kyr is
derived for the SNR, an order of magnitude larger than
the 1700~yr implied
by the pulsar spin-down (\cite{shmc83}). One possible
explanation for this discrepancy
is that the pulsar is much older than
it seems (\cite{br88}; \cite{gva01b}), but this requires
an unusual spin-down history of which there is no evidence
from long-term timing (\cite{kms+94}).
A more likely explanation is that the SNR has expanded rapidly
into a low-density cavity (\cite{shmc83}). This model
can also explain the unusual SNR morphology, the offset
of the pulsar from the SNR's apparent center, and the
faintness of any PWN at radio wavelengths (\cite{bha90};
G99). This argument is
supported by recent observations of \HI\ emission in the
region (\cite{dgg+02}).

X-ray emission from this system, also displayed in 
Figure~\ref{fig_g320_most_rosat}, has revealed a different
but similarly complicated picture. 
Apart from a pulsed point-source corresponding to
the pulsar itself (\cite{sh82}), X-ray emission from the
system is dominated by an elongated non-thermal source centered on the
pulsar, and an extended thermal source coincident with the northern radio
component of the SNR and with the optical nebula \rcw\ (\cite{shmc83};
\cite{tmc+96}). Faint diffuse X-rays extend over the entire
extent of the radio SNR, and potentially correspond to thermal
emission from the SNR blast wave (\cite{tmc+96}).

The central 
non-thermal source has been interpreted as a PWN powered by the pulsar
(\cite{shss84}; \cite{tmc+96}).  As mentioned above, no 
radio emission from this PWN has been seen, presumably
because of the low density into which the pulsar wind 
expands (\cite{bha90}; G99).  High-resolution observations with
{\em ROSAT}\ show the X-ray PWN to be significantly collimated to the
SE of the pulsar (Fig.~\ref{fig_g320_most_rosat}; 
\cite{gcm+95}). Several authors have interpreted
this region as corresponding to a jet produced by the pulsar
(\cite{tkyb96}; \cite{bb97}).

Using {\em ROSAT}\ PSPC data, Greiveldinger
\etal\ (1995\nocite{gcm+95}) have analyzed the morphology of X-ray
emission close to the pulsar, and concluded that the PWN has two
components: a large region extending over many arcminutes, and
a compact disc of emission of radius $30''-45''$ centered on the
pulsar.  Brazier \& Becker (1997\nocite{bb97}) considered data taken
with the {\em ROSAT}\ HRI. They found no evidence for the two-component
PWN proposed by Greiveldinger \etal\ (1995\nocite{gcm+95}), but pointed
out that the X-ray emission around the pulsar had a cross-shaped
morphology. They interpreted this as indicating that the pulsar powers
an equatorial torus, as is also seen in X-rays for the Crab 
pulsar (\cite{ab75}; \cite{hss+95}).  This
morphology would imply that the jet (and presumably the pulsar
spin axis) lies largely in the plane of the sky, inclined at
$\sim70^\circ$ from the line-of-sight.

An interpretation for the thermal X-ray source $\sim6'$ to the north of
the pulsar is unclear. It has been repeatedly proposed that this region
represents an interaction between a collimated outflow from the pulsar
and the surrounding SNR (\cite{shmc83}; \cite{md83}; \cite{man87};
\cite{bb97}). Tamura \etal\ (1996\nocite{tkyb96}) showed evidence for a
bridge of non-thermal emission connecting the pulsar to the thermal
X-ray emission to the north, but it remained uncertain 
as to whether this was the
direct counterpart to the SE jet-like feature.  At high spatial
resolution, a collection of X-ray clumps is seen at the center of the
northern X-ray region, the morphologies and positions of which closely match
those of clumps seen at radio wavelengths (\cite{bb97}; G99). It
has not been established whether the proposed pulsar outflow powers the
entire \rcw\ region (\cite{tkyb96}) or just the central collection of
clumps (G99).  But in either case, the evidence for interaction
provides a convincing argument that the pulsar and SNR are physically
associated.

Clearly PSR~\psr\ and SNR~\snr\ provide an important example of the
interaction between a pulsar and its environment. Accordingly, we have
obtained sensitive high-resolution observations of this system with the
{\em Chandra X-ray Observatory}. These data provide the opportunity to
clarify the nature of the emission close to the pulsar, to 
determine the overall structure of the PWN, and to further investigate
the apparent interaction between the pulsar and its surrounding SNR. In
\S\ref{sec_obs} we describe our observations, in \S\ref{sec_results} we
describe the resulting images and spectra, and in \S\ref{sec_discuss}
we discuss and interpret these results.

\section{Observations and Analysis}
\label{sec_obs}

\snr\ was observed on 2000~Aug~14 (ObsID 754)
with the Advanced CCD Imaging Spectrometer (ACIS)
aboard the {\em Chandra X-ray Observatory}\ (\cite{wtvo00}).
ACIS is an array of ten 1024$\times$1024-pixel  CCDs
fabricated by MIT Lincoln Laboratory.   These X-ray sensitive devices
have large detection efficiency (10$-$90\%) and moderate energy
resolution (10$-$50) over a 0.2--10.0~keV passband.  Coupled with the
10-m optics of \cxo, the CCDs have a scale of $0\farcs492$ per pixel,
well-matched to the on-axis point-spread function (FWHM $\la1''$).
ACIS and its calibration are discussed in detail by Burke \etal\
(1997\nocite{bgb+97}) and Bautz \etal\ (1998\nocite{bpb+98}).

A single exposure of length 20~ks was made in the standard ``Timed
Exposure'' mode. The data discussed here were recorded from the four
CCDs of the ACIS-I array.  After standard processing had been carried
out at the Chandra Science Center (ASCDS version number R4CU5UPD14.1),
we analyzed the resultant events list using CIAO v2.1.3. We first
applied the observation-specific list of bad pixels to the data, and
then corrected for the $\sim1''$ offset present in ACIS-I
observations\footnote{\tt
http://asc.harvard.edu/mta/ASPECT/fix\_offset.cgi}.  After removing
small intervals in which data were not recorded, the final exposure
time for this observation was 19039~s.

The effective area of the detector is a function of 
both the energy of incident photons and their position
on the sky. To generate an exposure-corrected image,
we computed exposure maps at five
energies: 0.5, 1.5, 3.0, 5.0 and 7.0~keV.
We also computed images of the counts per pixel
in the five separate energy bands 0.3--0.8, 0.8--2.0,
2.0--4.0, 4.0--6.0 and 6.0--8.0~keV. For each energy
band, we divided the map of raw counts by the exposure map corresponding
to that band, and then normalized by the exposure time
to produce a final image with units of photons~cm$^{-2}$~s$^{-1}$. 

For compact sources of emission ($\le5''$ in extent), spectra were
extracted from the data using the {\tt psextract}\ script within CIAO,
using a small extraction region immediately surrounding the source of
interest.  For extended sources we made use of v1.08 of the {\tt
calcrmf} and {\tt calcarf} routines\footnote{\tt
http://asc.harvard.edu/cont-soft/software/calcrmf.1.08.html} provided
by Alexey Vikhlinin, which generate the response matrix and effective
area for a region of the CCD by computing an appropriately weighted
mean over small sub-regions.  In all cases, the resulting spectra were
rebinned to ensure a minimum of 15 counts per spectral channel.  The
background spectrum for each region was measured by considering an
annular region surrounding the source of interest.  Subsequent spectral
analysis was carried out using XSPEC v11.0.1.

\section{Results}
\label{sec_results}

\subsection{Imaging}

\begin{figure*}[hbt]
\vspace{1cm}
\centerline{\psfig{file=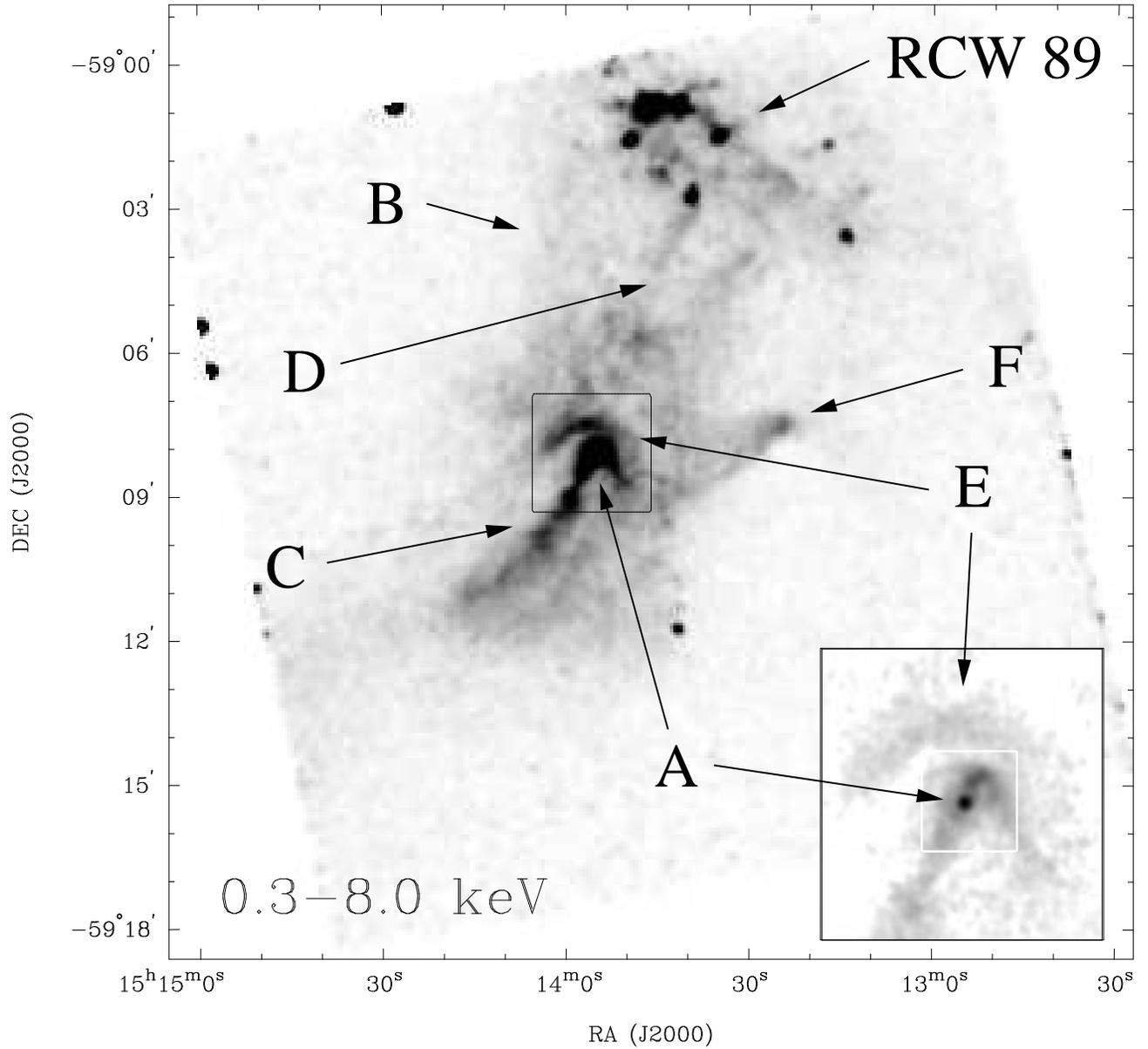,height=16cm}}
\caption{\protect\cxo\ image of \snr\ over the entire
energy band. The image has been
exposure-corrected and convolved with a gaussian
of FWHM $10''$, and is displayed using a linear transfer function.
The central black box in the main image
shows the region covered by the panel at lower
right. In this panel, the 
data have been smoothed with a gaussian of FWHM $4''$ and displayed using
a logarithmic scale; the white box in this panel
shows the region displayed in Figure~\ref{fig_g320_center}.
Specific features discussed in the text are indicated.}
\label{fig_g320_acisi}
\end{figure*}

An exposure-corrected and smoothed image in the energy range
0.3--8.0~keV, encompassing the entire field-of-view of the ACIS-I
array, is shown in Figure~\ref{fig_g320_acisi}. The central region is
shown on a logarithmic scale in the subpanel at lower-right.  Count
rates and surface brightnesses for various features (marked in
Fig.~\ref{fig_g320_acisi} and discussed below) are given in
Table~\ref{tab_rates}.

Marked in Figure~\ref{fig_g320_acisi} as feature~A,
PSR~\psr\ is clearly detected as a bright source near the center
of the field-of-view. The extent of this source is consistent
with the point-spread function of the
telescope. The pulsar's X-ray position
is measured to be
(J2000) RA~$15^{\rm h}13^{\rm m}55\farcs64$,
Dec~$-59^\circ08'09\farcs2$, with an
uncertainty of $\pm0\farcs5$ in each coordinate.
This differs by several arcsec from previous
X-ray measurements of the pulsar position (\cite{shss84}; \cite{bb97}),
but is in excellent agreement with positions measured
from radio timing and imaging (\cite{kms+94}; G99\nocite{gbm+98}).

Because the pulsar is so bright, a significant number
of counts from it strike the detector during the
41~ms per 3.2-sec frame during which the CCD is read out. This
produces a faint trail of emission,
marked as feature~B, which runs through the
pulsar and parallel to the north/south edge of the array (i.e.\
along the read-out direction).

The morphology of the surrounding X-ray emission
as imaged with \cxo\ is consistent with that seen by {\em ROSAT}\ 
and {\em ASCA}\ at
lower spatial resolution.  In particular, we confirm the diffuse nebula
seen extending $\sim6'$ NW/SE of the pulsar (\cite{gcm+95};
\cite{tmc+96}; \cite{tkyb96}), and the collection of X-ray clumps coincident
with the H$\alpha$ nebula 
RCW~89 (\cite{bb97}).  Most of the nebula is elongated along a 
position angle $150^\circ\pm5^\circ$ (N through E). In all
future discussion, we consider this orientation to define
the main axis of the system.

To the SE of the pulsar and superimposed on the diffuse
nebular emission is a jet-like X-ray feature, denoted
as feature~C in Figure~\ref{fig_g320_acisi}. Close to
the pulsar, feature~C is aligned with the
main nebular axis, but at a distance $\sim2'$ from the
pulsar it curves around to the east, before fading
abruptly at a separation of
$\sim4'$ from the pulsar. This feature is resolved and
is approximately $10''$ wide. 

\begin{figure*}[hbt]
\vspace{1cm}
\centerline{\psfig{file=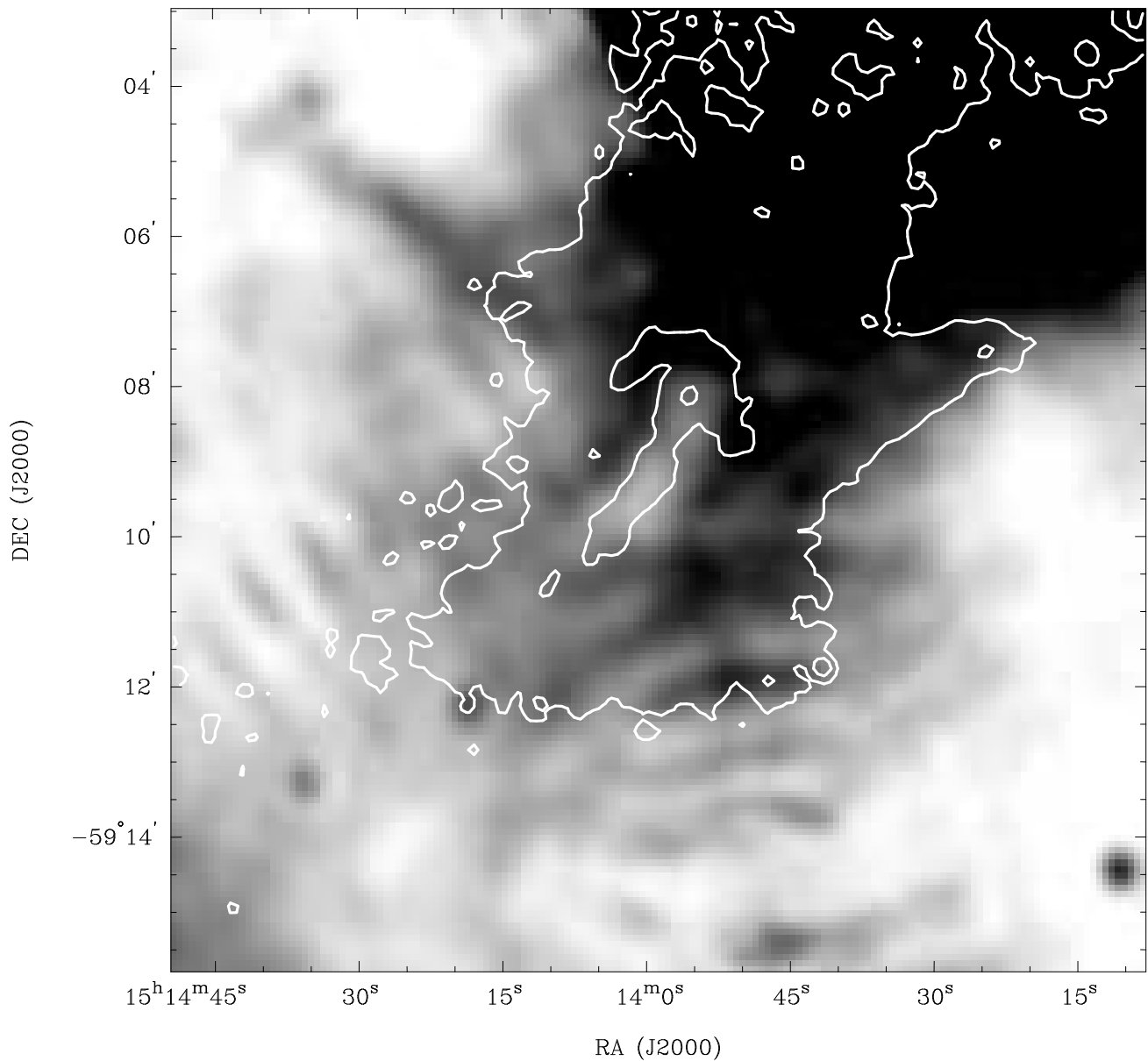,height=16cm,angle=270}}
\caption{A comparison of X-ray and radio emission
in the region surrounding PSR~\psr. The greyscale displays radio data 
from 1.4-GHz observations with the ATCA
at a spatial resolution of $21''\times24''$
and over a range 0 to 10~mJy~beam$^{-1}$, as presented by G99.
The ripples seen in the radio emission are residual
sidelobes from the bright \rcw\ region to the north.
The contours show \cxo\ data as in Figure~\ref{fig_g320_acisi},
with contours at 0.5\%, 2\% and 30\% of the peak.}
\label{fig_g320_tongue}
\end{figure*}

In Figure~\ref{fig_g320_tongue} we show a comparison between our \cxo\
data and 1.4-GHz radio observations taken with the Australia
Telescope Compact Array (ATCA; G99).
There is a good match between the morphologies
of the X-ray PWN and the radio ``tongue'' reported
in previous observations (\cite{md83}; G99), 
the radio extent being somewhat larger than the X-ray nebula.
Furthermore, it is clear that feature~C corresponds
to a trough of reduced radio emission, confirming
a similar correspondence seen when comparing
ATCA and {\em ROSAT} observations (G99).

We see no obvious counterpart in Figure~\ref{fig_g320_acisi}
to feature~C on the other
side of the pulsar. Rather, to
the NW of the pulsar there is
an elongated region of {\em reduced}\ emission, beginning $3'$
from the pulsar and extending for $3'$ along the nebular axis.
We refer to this as feature~D.  The
width of this fainter region is a function of distance from the pulsar:
at its end nearest the pulsar, this feature is $\sim25''$ across, but it
widens to $75''$ at its NW end. 

Approximately $30''$ to the north of the pulsar is feature~E,
a semi-circular arc
of emission, approximately centered on the pulsar. This 
arc is resolved by \cxo, with approximate width $15''$. 
The arc is brightest immediately to the north of the pulsar, fading
slowly to the east and more rapidly to the west. Along the
main symmetry axis defined above, the pulsar is separated from the
inner edge of this arc by $30''$, and from the
outer edge by $45''$. 

\begin{figure*}[hbt]
\vspace{1cm}
\centerline{\psfig{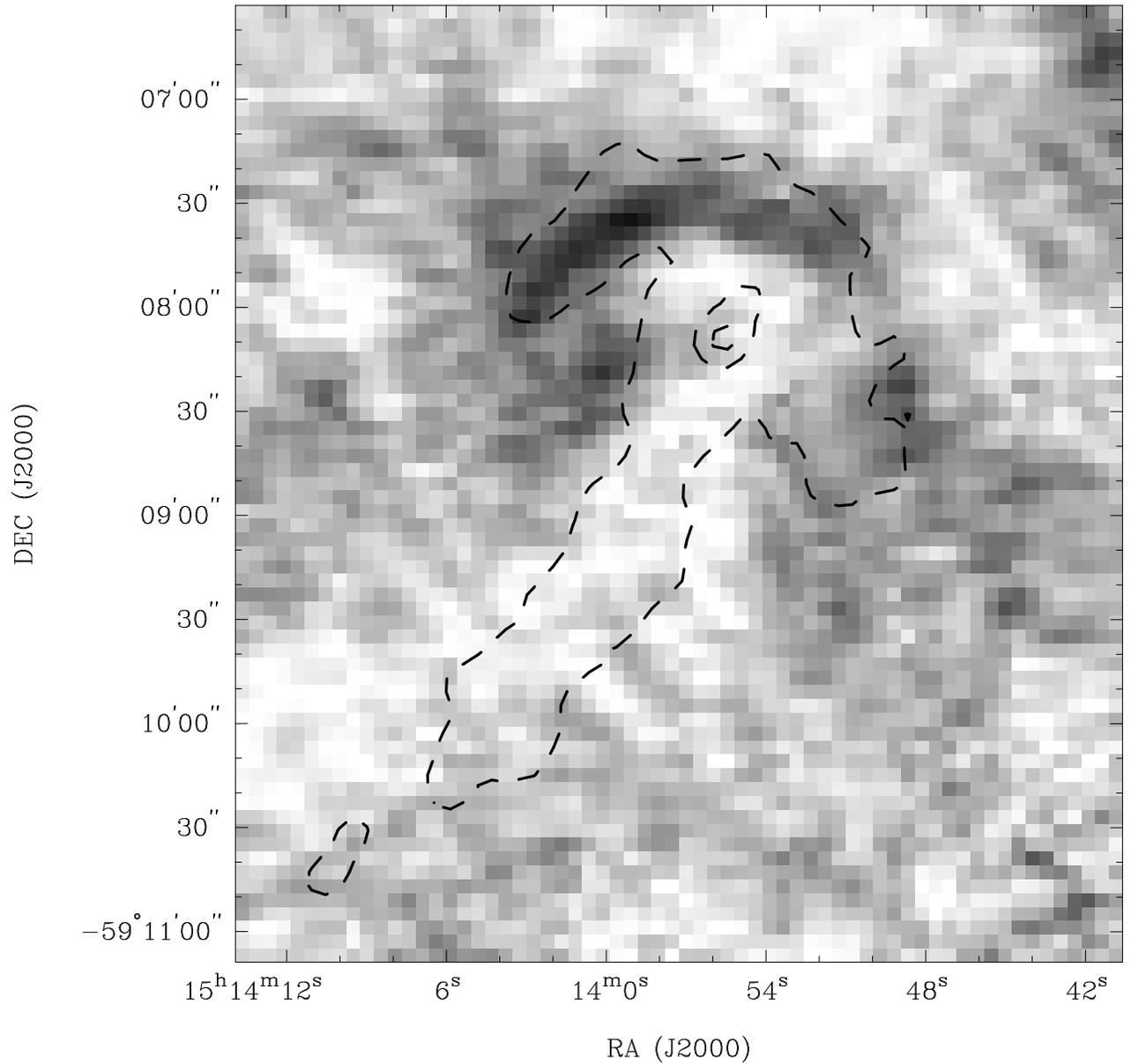}}
\caption{Polarized radio emission in the region near the pulsar.
The greyscale shows linearly polarized intensity 
at  5~GHz, at a spatial resolution of
$6''\times9''$ and over a range of 0 to 0.85~mJy~beam$^{-1}$.
The contours show \cxo\ data as in Figure~\ref{fig_g320_acisi}, with
contours at 2\%, 20\% and 50\% of the peak.}
\label{fig_g320_pol}
\end{figure*}

We have re-examined the 1.4- and 5-GHz images presented by G99 
to look for a radio counterpart to feature~E. 
In total intensity, this
region is confused by emission from \rcw\ (see
Fig.~\ref{fig_g320_tongue}), and we can identify no specific radio
feature which might be associated.  However, we have
re-analyzed the 5-GHz polarization data of G99, deconvolving the data
with a new maximum entropy algorithm specifically designed for
polarimetric observations ({\tt PMOSMEM}; \cite{sbd99}). In
Figure~\ref{fig_g320_pol} we compare the resulting distribution of linear
polarization  to the X-ray emission in the same region.  While there is
generally a high level of polarization from the area, there is a
clear enhancement of the linearly polarized signal in an arc $\sim30''$
to the north of the pulsar, whose extent and morphology match closely
that of the X-ray emission from feature~E.

Returning to Figure~\ref{fig_g320_acisi}, it can be seen that
the perimeter of the overall diffuse nebula is reasonably
well-defined. This is particularly the case to the west of the pulsar,
where feature~F, a sharp-edged protuberance, extends $\sim2'$ to the
WNW of the main body of the nebula. A similar, but smaller and less
well-defined feature may exist on the opposite side of the nebular
symmetry axis.

\begin{figure*}[hbt]
\vspace{1cm}
\centerline{\psfig{file=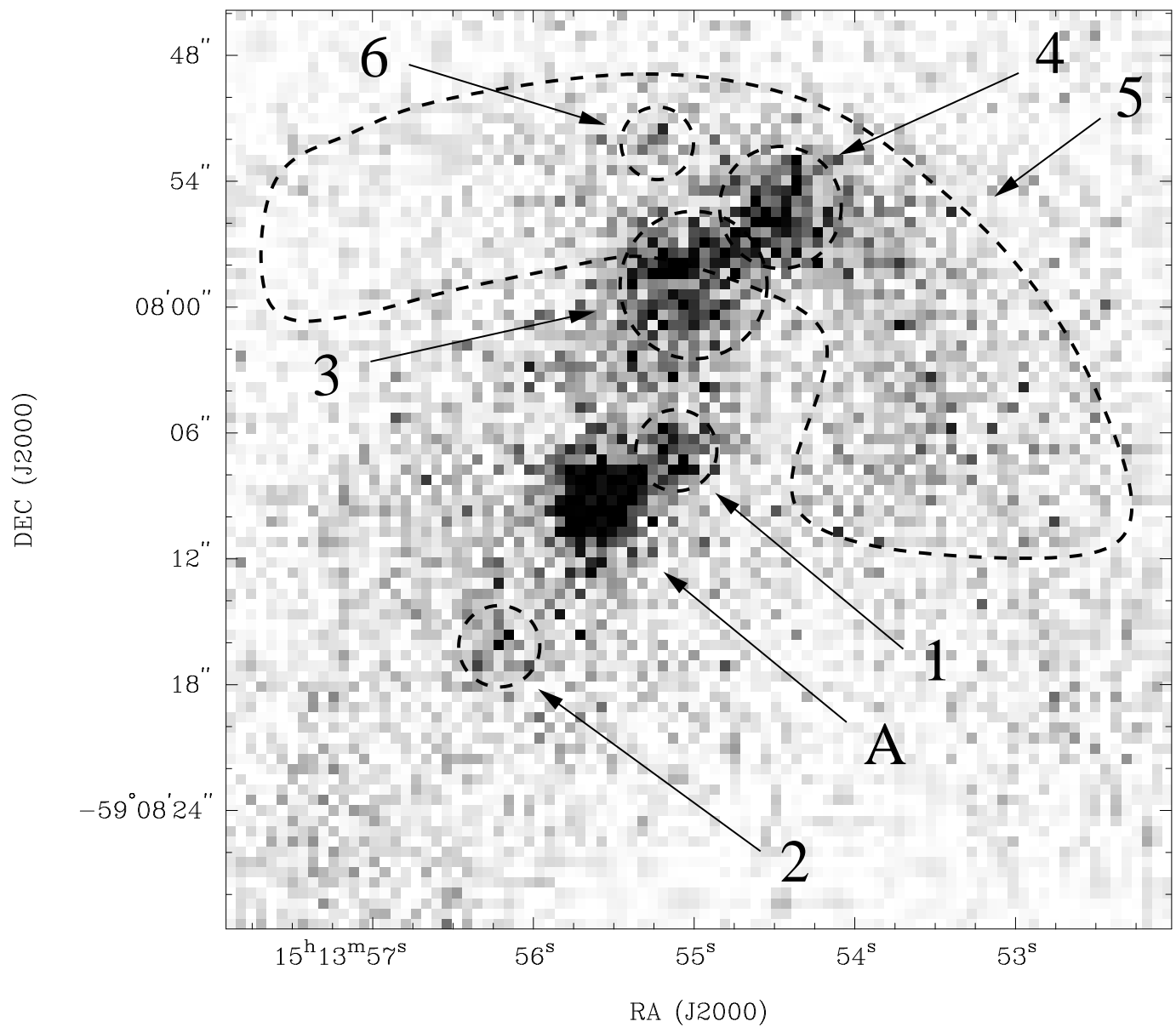,height=16cm}}
\caption{X-ray emission immediately surrounding PSR~\psr\ in
the energy range 0.3--8.0~keV.
These data have been exposure-corrected but have had no
smoothing applied. Specific features discussed in the text are marked
and labeled.}
\label{fig_g320_center}
\end{figure*}

Figure~\ref{fig_g320_center} shows the X-ray emission immediately
surrounding the pulsar at the full resolution of the data.
In increasing order of distance from
the pulsar (marked as A), we point out
various features of interest:

\begin{enumerate}

\item Approximately $3''$ to the NW of the pulsar, a small
clump of emission $\sim3''$ across. This feature is too
far from the pulsar and too irregular in appearance
to be due to asymmetries in the wings 
of the point-spread function (cf.\ \cite{pksg01}). 
The count rate from this source is
approximately constant over the observation, so it cannot
result from pulsar photons
assigned to the wrong location on the sky (as
might result from errors in the aspect solution).

\item  Lying $8''$ to the SE of the pulsar
along the main nebular axis, another small clump of emission.

\item Approximately $10''$ to the north
of the pulsar, a circular clump of diameter $3''$.

\item Immediately to the NW of feature~3, a similarly-sized
clump.

\item Superimposed on feature~4, a faint circular arc 
of emission. This arc is approximately centered on
the pulsar, with a radius of $17''$ and an angle
subtended at the pulsar of $\sim110^\circ$. The
midpoint of this feature lies
along the main nebular axis.

\item A faint unresolved source, positionally coincident  
with the O6.5III star Muzzio~10 (\cite{om80}; \cite{shmc83}).
\end{enumerate}

\subsection{Spectroscopy}

\begin{figure*}[hbt]
\vspace{1cm}
\centerline{\psfig{file=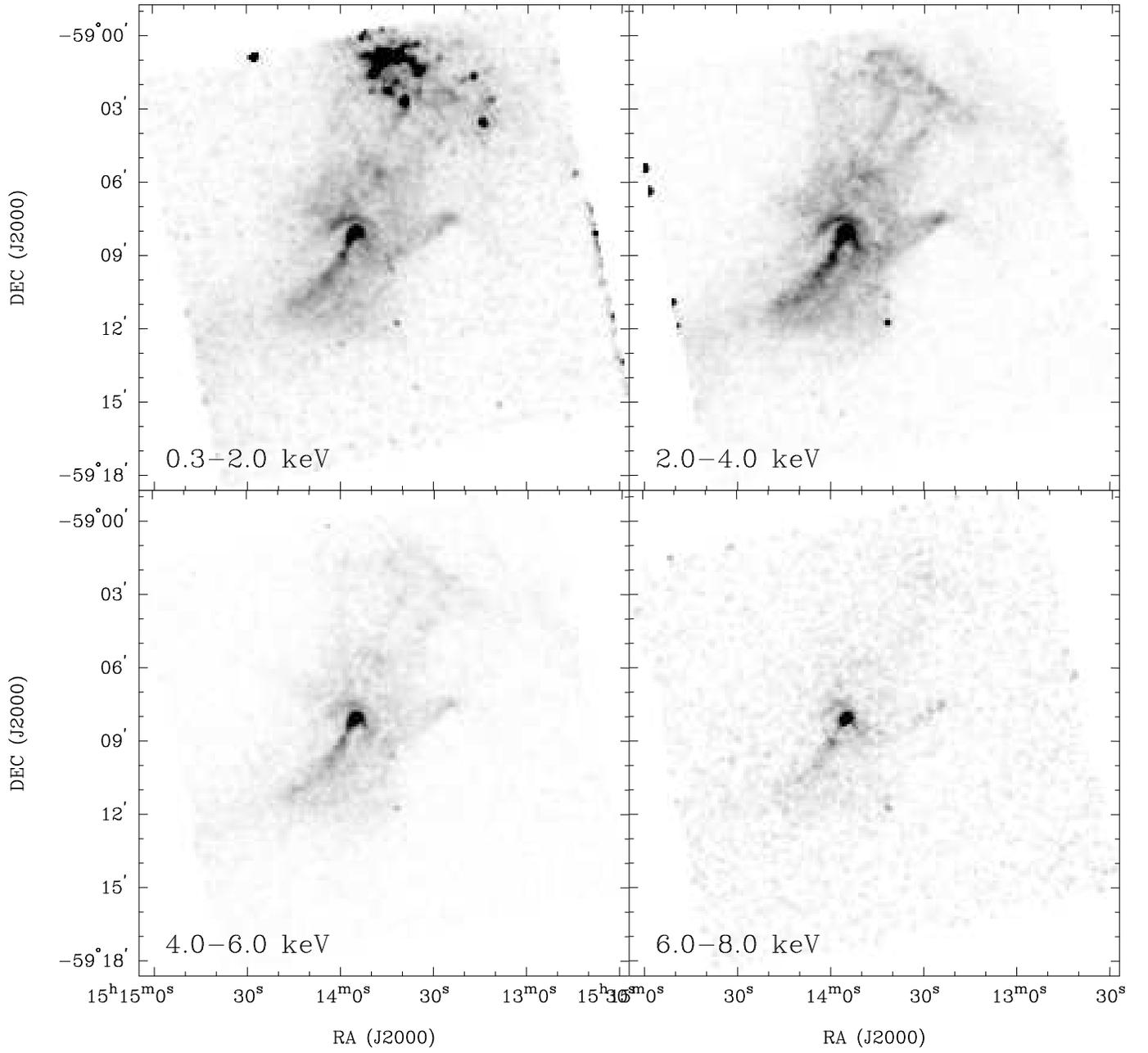,height=17cm,angle=270}}
\caption{As in Figure~\protect\ref{fig_g320_acisi}, but for
four energy sub-bands.}
\label{fig_g320_bands}
\end{figure*}

The gross spectral characteristics of the system can be derived from
Figure~\ref{fig_g320_bands}, where we show the field surrounding
SNR~\snr\ in four separate energy sub-bands.  This set of images
demonstrates that the northern ring of X-ray clumps coincident with
\rcw\ have no detectable emission above $\sim$4~keV, while the PWN has
significantly harder emission. The pulsar is clearly the hardest source
in the field, although this may largely be due to pile-up in its
spectrum (see further discussion below). These broad conclusions
confirm the spectral decompositions made from {\em Einstein} and {\em
ASCA}\ observations (\cite{shmc83}; \cite{tkyb96}).

At the higher spatial resolution of the \cxo\ data, some new spectral
features become apparent. The 2.0--4.0~keV and 4.0--6.0~keV images in
Figure~\ref{fig_g320_bands} demonstrate that the  PWN has an extended
component coincident with the \rcw\ clumps.  Furthermore, there is the
suggestion from Figure~\ref{fig_g320_bands} that the jet
and outer arc (features~C and E respectively) have
harder spectra than the overall PWN.

Spectral fits show that the point-source 
seen at RA~$15^{\rm h}13^{\rm m}41^{\rm s}$, Dec~$-59^{\circ}11'45''$
is heavily absorbed, with $N_H \sim 3\times10^{22}$~cm$^{-2}$ for
simple thermal and non-thermal models.
It therefore most likely represents an unrelated background source.

\subsubsection{PSR~\psr}

The pulsar itself has a high X-ray flux and is expected to suffer from
significant pile-up, in which multiple events arrive on a given CCD
pixel during a single frame.  Indeed the spectrum of the pulsar has an
excess of hard photons (presumably due to several lower-energy
photons being recorded as a single event), and cannot be fit by any
simple power-law model.  Assuming a foreground absorbing column $N_H
\sim 1\times10^{22}$~cm$^{-2}$, 
a power-law spectrum with photon
index $\Gamma  \sim 1.4$, and an unabsorbed flux $f_x \sim
6\times10^{-12}$~erg~cm$^{-2}$~s$^{-1}$ in the energy range
0.1--2.4~keV (\cite{gcm+95}), the \cxo\ proposal planning
software\footnote{{\tt http://asc.harvard.edu/toolkit/pimms.jsp}}
predicts $>$50\% pile-up, with a resulting detected count rate 0.16
cts~s$^{-1}$ over the energy
range 0.5--10~keV.  This in reasonable agreement with the value measured
for the pulsar in Table~\ref{tab_rates}.

\subsubsection{The Pulsar Wind Nebula}
\label{sec_spec_pwn}

\begin{figure*}[hbt]
\vspace{1cm}
\centerline{\psfig{file=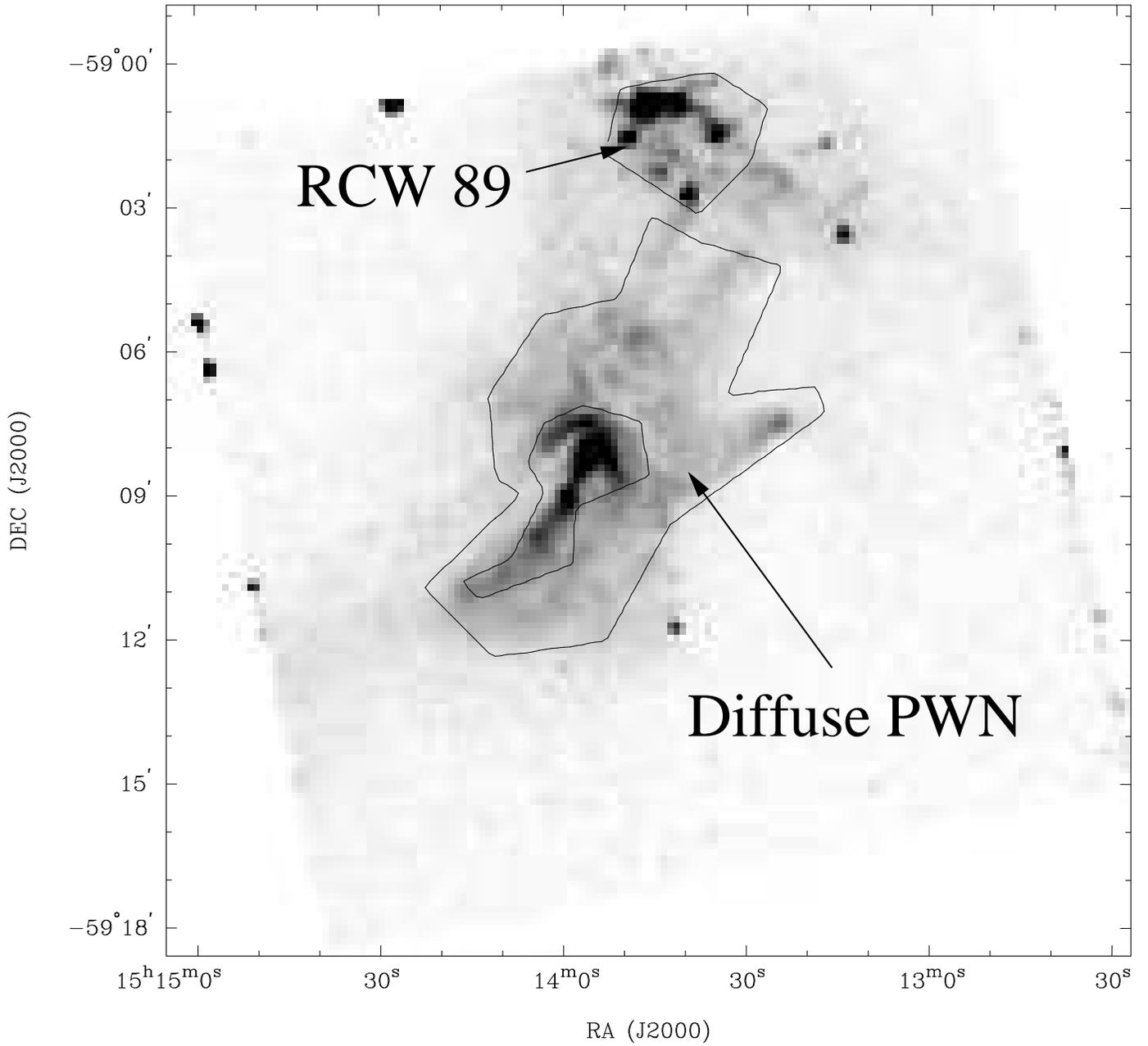,height=17cm}}
\caption{A low-resolution version of Figure~\ref{fig_g320_acisi},
showing the extraction regions used to generate
the spectra shown in Figures~\ref{fig_pwn_spec} and \ref{fig_clump_spec}.
The region marked ``Diffuse PWN'' is an annular region which
excludes regions A, C and E and features 1--5.}
\label{fig_g320_regions}
\end{figure*}

\begin{figure*}[hbt]
\vspace{1cm}
\centerline{\psfig{file=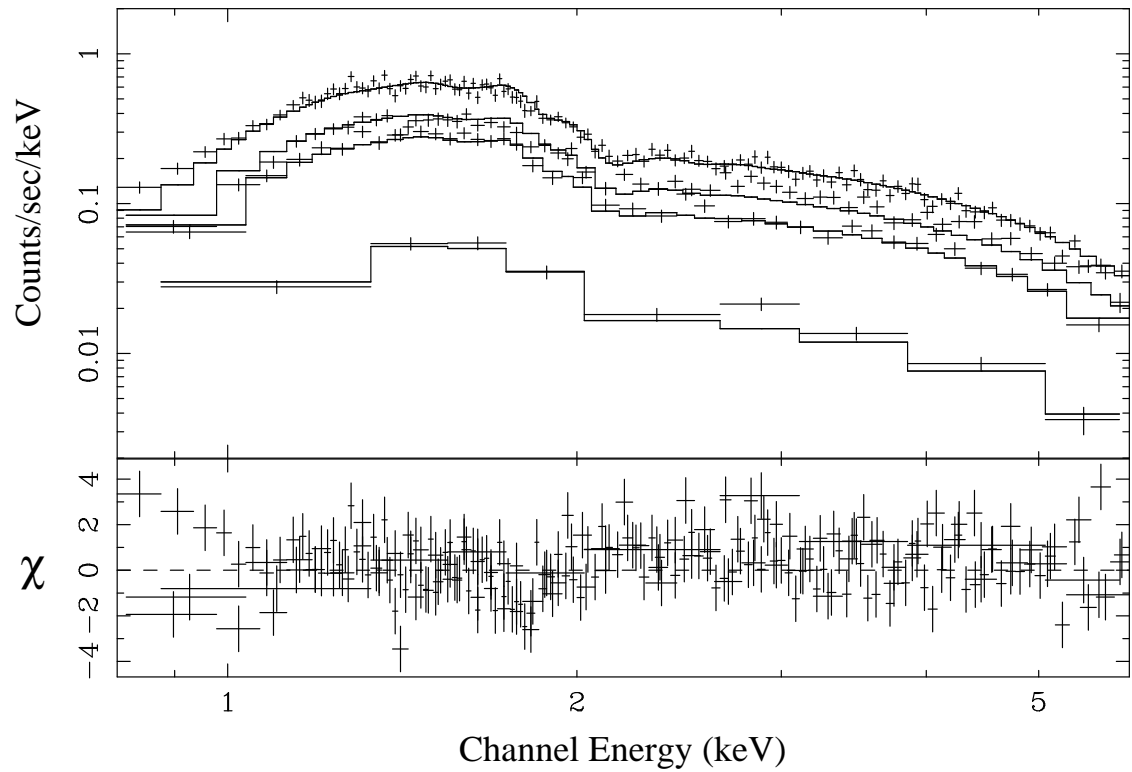,height=12cm}}
\caption{X-ray spectrum of the diffuse pulsar wind nebula.
The data-points in the upper panel show the data from each of the four
CCDs on which the nebular emission falls, while
the solid lines correspond to the best-fit absorbed power-law model. 
The lower panel shows the number of standard deviations
by which the model and the data differ in each bin.
The data have been plotted so as to give a signal-to-noise ratio
of 10 in each bin.}
\label{fig_pwn_spec}
\end{figure*}

We can determine the spectrum for the diffuse nebula
surrounding the pulsar by extracting photon energies
from the annular region shown in Figure~\ref{fig_g320_regions}.
Measuring the spectrum for 
an extended source such as this is difficult because
of radiation
damage to the front-illuminated 
CCDs,\footnote{\tt http://asc.harvard.edu/cal/Links/Acis/acis/Cal\_prods/qe/qe\_warning.html}, which has caused the response of each CCD
to be a function of distance from the read-out nodes.
To try and mitigate this effect, we extracted
separate spectra for each of the four CCDs on which this diffuse emission
falls, and generated response matrices and effective
area curves for each CCD separately.
We then fit these four spectra simultaneously over
the entire usable energy range (0.5--10.0~keV), using
a common value for all fit parameters except the normalization
of each spectrum.
These spectra, which between them represent a total of
$\sim$50\,000 photons, are shown
in Figure~\ref{fig_pwn_spec}. While no significant
spectral features are seen, some systematic residuals are still
present, which we attribute to gain mismatches
resulting from the aforementioned radiation damage.
The count rate from the diffuse nebula is very high
($\sim$3~cts~s$^{-1}$), but this emission is spread over
a very large area so that pile-up is negligible. 

We find that these data are well fitted by an
absorbed power law. As listed
in Table~\ref{tab_spec}, the best-fit
spectral parameters are an
absorbing column $N_H = (9.5\pm0.3)\times10^{21}$~cm$^{-2}$
and a photon index $\Gamma =2.05\pm0.04$.

We next consider the spectra of each of regions~C and E and of features
1--5.  All these sources have a surface brightness far too low to
produce pile-up in the detectors.  We fit absorbed power laws to the
data in each region over 0.5--10.0~keV, with the resulting spectral
parameters listed in Table~\ref{tab_spec}.  With the exception of
feature~2 (which  has very few counts and correspondingly uncertain
spectral parameters), the fitted values of $N_H$ are all consistent
with each other and with that for the diffuse PWN,
while the resulting photon indices range between
$\Gamma \approx 1.3$ and $\Gamma \approx 1.7$. These
photon indices are all significantly
flatter than that determined for the diffuse PWN.  To more tightly
constrain the photon index in these regions, we assume that they all
have the same absorbing column as that determined for the diffuse
nebula above, and thus fix $N_H$ to $9.5\times10^{21}$~cm$^{-2}$
corresponding to the power law fit to that source.  The
resulting spectral fits confirm the flatter spectra for these regions
when compared to the overall nebula.

Finally, we crudely fit a Raymond-Smith spectrum
to the emission from feature~6, coincident with 
the star Muzzio~10. The approximate spectral
parameters are listed in Table~\ref{tab_spec}.

\subsubsection{\rcw}

\begin{figure*}[hbt]
\vspace{1cm}
\centerline{\psfig{file=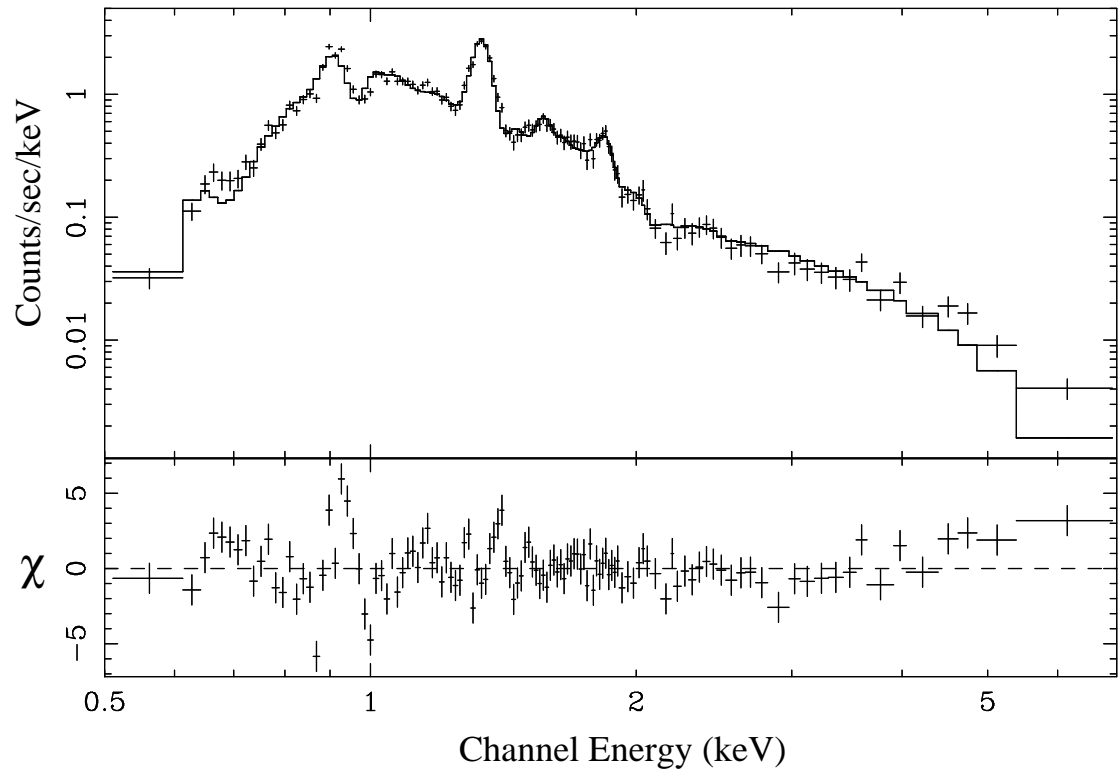,height=12cm}}
\caption{As in Figure~\ref{fig_pwn_spec},
but for a spectral fit  to 
the brightest clumps in \rcw.  In this case, the 
solid line corresponds to a non-equilibrium ionization
model with variable abundances,
and the data have been plotted with a signal-to-noise
of 8 in each bin.}
\label{fig_clump_spec}
\end{figure*}

The spectrum for the \rcw\ region is considerably more complex than for
the PWN.  In Figure~\ref{fig_clump_spec} we show a spectrum for the
region shown in Figure~\ref{fig_g320_regions}, enclosing the brightest
clumps in \rcw\ (sources N0, N1, N2, N3, N5, N6, N7 and N8 in the
designation of \cite{bb97}).  This spectrum is clearly
dominated by emission lines. We also show a crude fit to the data
using a non-equilibrium ionization model with variable abundances
(model ``{\tt vnei}'' in XSPEC).  The fit is poor
($\chi_\nu^2/\nu=455/254=1.91$), but this is mainly due to large
residuals in the Ne emission line at 0.9~keV.  In particular, the
continuum component of the spectrum of \rcw\ is well-accounted for,
with little emission above 3--4~keV.

While we defer a full and detailed treatment of these data to a
subsequent paper, we think it clear from Figures~\ref{fig_g320_bands} and
\ref{fig_clump_spec} that we can rule out the possibility
that \rcw\ consists of synchrotron-emitting clumps embedded in a diffuse
thermal nebulae (as argued by G99), and that a model involving thermal
clumps embedded in a diffuse synchrotron nebula (as proposed by
\cite{tkyb96}) seems far more likely.

\section{Discussion}
\label{sec_discuss}

\subsection{The Overall Nebula}
\label{sec_discuss_pwn}

It is immediately clear from the images presented here that
the PWN surrounding PSR~\psr\ has a very complicated
morphology, matched only by the PWNe powered by the Crab
and Vela pulsars in the level of structure. 

We first note that there is a clear axis of symmetry associated
with the system. This is manifested not only in the overall
elongation of the nebula, but also in the orientation
of features C, D and E in Figure~\ref{fig_g320_acisi},
and features 1, 2, 4 and 5 in Figure~\ref{fig_g320_center}.
This axial structure is manifested on an extremely
wide range of scales, from $10'' = 0.2$~pc up
to $10' = 15$~pc (where here and in further discussion
we adopt a distance to the system of 5~kpc). It is hard to see how this
orientation could be due to interaction with
a pre-existing structure in the ISM (e.g.\ \cite{cas79}), which should
generally only influence the overall morphology
of the system but not its small-scale structure.
We therefore believe that this overall symmetry
reflects the properties of the central pulsar.

The only well-defined axes associated with a neutron
star are its spin axis and its velocity vector.
PWNe whose morphologies are dominated
by the pulsar's motion are axially symmetric but inevitably
show a clear cometary morphology with the pulsar at
one end (\cite{fggd96}; \cite{ocw+01}; \cite{kggl01}). In contrast, 
the PWN surrounding PSR~\psr\ extends to many parsecs
on both sides of the pulsar. It is thus clear that
any motion of the pulsar cannot explain the
overall PWN morphology (see additional discussion
in \S\ref{disc_bowshock}), and
that the main axis of the nebula must correspond
to the pulsar spin axis. Such a correspondence
was previously proposed for this pulsar by Brazier
\& Becker (1997\nocite{bb97}), 
has also been argued for the Crab and Vela
pulsars and their nebulae (\cite{hss+95}; \cite{psg+00}; \cite{hgh01}),
and is seen in some models for
pulsar magnetospheres (\cite{sl90}). 

The region of the PWN where the symmetry of the system is most clearly
broken is in the vicinity of feature~F, the promontory extending
several arcmin west of the main axis. The sharp edges seen for this
feature are reminiscent of the ``bay'' observed along the western edge of the
Crab Nebula  (\cite{wht+00}).  Such morphologies are presumably
produced by strong confining pressure in these regions, possibly the
result of pre-existing magnetic fields or circumstellar material.

The spectral fits in Table~\ref{tab_spec} show that
X-ray emission from the diffuse component of the PWN can
be well fit by a power-law with
photon index $\Gamma = 2.05\pm0.04$ and absorbing
column $(9.5\pm0.3)\times10^{21}$~cm$^{-2}$. This
measurement is in good agreement with
previous spectral measurements 
for this source (\cite{kos93};
\cite{tmc+96}; \cite{tkyb96}; \cite{mbg+97}; \cite{mcm+01}).

Du Plessis \etal\ (1995\nocite{ddb+95}) have argued from a collection
of archival X-ray data sets that the photon index for this PWN steepens
from $\Gamma \approx 1.6$ below $\varepsilon \sim 6$~keV to $\Gamma \approx
2.15$ above this energy. We see no evidence for this spectral curvature
in our data set: a fit to the data using a broken power-law model rules
out any spectral break across the \cxo\ bandpass larger than $\Delta
\Gamma \ga 0.2$. It is more likely that the effect 
claimed by du Plessis \etal\ (1995\nocite{ddb+95}) was a result
of the large uncertainties in and calibration differences
between their lower sensitivity data sets.

The standard observational picture for a PWN is that its
spectrum is comparatively flat
at radio wavelengths, $1 \la \Gamma \la 1.3$, but
is steeper at X-ray energies, typically  with
$\Gamma \approx 2$ (see \cite{gae01b}
and references therein). It is usually assumed
that synchrotron losses are at least partly
responsible for this steepening, generating
a spectral break
at energy $\varepsilon_s$, across which we expect a change in 
photon index $\Delta\Gamma = 0.5$ (\cite{kar62}; \cite{ps73}).
The frequency of this break, along with an
estimate of the age of the PWN, can be used to
infer the nebular magnetic field (e.g.\ \cite{fggd96}). In the
case of the PWN around PSR~\psr,
the observed photon index $\Gamma = 2.05$
implies $\varepsilon_s \la 1$~keV.
If the system is $t$~yr old, then the implied
nebular magnetic field is
\begin{equation}
B_n = 8.0\left(\frac{t}{1700~{\rm yr}}\right)^{-2/3}
\left(\frac{\varepsilon_s}{1~{\rm keV}}\right)^{-1/3}~\mu{\rm G},
\label{eqn_b}
\end{equation}
so that we can infer $B_n \ga 8$~$\mu$G.

There are a number of other, independent, estimates
of the nebular magnetic field: the assumption
of equipartition (\cite{shss84}), 
a simple model of PWN evolution (\cite{shss84}) and
a detection of inverse-Compton emission from the PWN
at TeV energies (\cite{smm+00b}) all
result in estimates in the range $B_n = 5-8$~$\mu$G,
while Chevalier (2000\nocite{che00}) points out
that the comparatively low efficiency with which  this pulsar
converts its spin-down energy into X-ray synchrotron emission
also implies $\varepsilon_s \la 4$~keV and hence
$B_n \ga 5$~$\mu$G. 

We thus conclude from a variety
of arguments that the synchrotron break in this PWN
is just below the X-ray band, and that
the magnetic field is correspondingly low, 
$B_n \approx$~8~$\mu$G (cf.\ $B_n \ga 100$~$\mu$G for the
Crab). This low magnetic field
presumably results from the low confining
pressure into which this PWN expands (\cite{shss84}).
It is worth noting that if we assume a pulsar age $t\sim20$~kyr (\cite{br88};
\cite{gva01b}), we infer from
Equation~(\ref{eqn_b}) that $B_n \sim 1.5$~$\mu$G, which is
at odds with all the other estimates described above.

The above arguments imply that the synchrotron lifetime of electrons
emitting in the \cxo\ band is similar to the age of the pulsar
and nebula. Thus while synchrotron-emitting electrons
in the outer diffuse nebula have now radiated
most of their energy, we expect that regions closer to the pulsar
will have a flow time less than the synchrotron lifetime and
thus should have a spectrum flatter by $\Delta\Gamma 
\sim 0.5$.\footnote{In fact the magnetic field in the inner
parts of the PWN will generally be lower than the nebular
average (\S\ref{sec_wisps_struc} below; \cite{kc84a}), increasing the
synchrotron lifetimes in these inner regions even further.}
Indeed the various X-ray
features seen close to the pulsar (to be discussed in
more detail in subsequent sections) all have photon
indices $\Gamma \sim 1.5$.

Figure~\ref{fig_g320_tongue} demonstrates a good match between the
perimeter of the diffuse X-ray PWN and the ``tongue'' of emission
apparent in radio observations.  This correspondence was not apparent
in earlier observations, which lacked the sensitivity to delineate the
full extent of the X-ray PWN in this region (cf.\ Fig.~4 of G99). Given
that the two regions of emission occupy similar extents and have
similar shapes, we propose that the ``tongue'' simply corresponds to
emission from the pulsar nebula, and is the long-sought radio PWN in
this system. This region has previously been shown
to have a very high fractional polarization and a well-ordered
magnetic field (\cite{mch93}; G99), properties typical of a radio PWN.  

The radio PWN is comparatively faint, sits on a
complicated background, and is in close proximity to the very bright
radio emission from \rcw. The parameters of the
radio PWN are thus poorly constrained,
but using the data of Whiteoak \& Green
(1996\nocite{wg96}) and of G99, we estimate an approximate flux density
of $2\pm1$~Jy at both 0.8~GHz and 1.4~GHz. If we assume that this
represents $\sim$50\% of the total radio flux density of the PWN (the
rest being hidden by \rcw), we find $f_{1~{\rm GHz}} \sim
(1\pm0.5)\times 10^{-5}$~erg~s$^{-1}$~cm$^{-2}$~keV$^{-1}$. By
comparing this to the flux density for the diffuse 
X-ray nebula inferred from
Table~\ref{tab_spec}, 
$f_{1~{\rm keV}} = 1.9\times10^{-11}$~erg~~cm$^{-2}$~keV$^{-1}$,
we can infer that an unbroken power law with photon index $\Gamma
\approx 1.6$ is required to connect the two fluxes. This is
is comparable to
the photon index for the inner emitting regions in the nebula
which have not yet undergone synchrotron cooling, and
is consistent with the spectrum $1 < \Gamma < 2$ inferred for the radio
emission (``region 6'' in the analysis of G99). This
good agreement with these other estimates of
the uncooled synchrotron spectrum provides further
evidence that the spectrum of the PWN must steepen from $\Gamma \sim
1.5$ to $\Gamma \approx 2.05$ just below the X-ray band,
supporting the value $B_n \sim 8$~$\mu$G inferred above.

The radio observations shown in Figure~\ref{fig_g320_tongue}
were designed to study the overall structure of the
SNR, and were thus not optimized to map out the
fainter radio emission surrounding the pulsar. Further
radio observations are planned which we hope will
allow us to better delineate the emission from this region.

\subsection{Outflow and Pulsar Orientation}
\label{sec_discuss_orient}

Close to the pulsar, the collimated jet-like 
feature~C is narrow, bright, and closely aligned
with the symmetry axis; at larger distances it fades, broadens
and gradually changes direction. Furthermore, Figure~\ref{fig_g320_pol}
demonstrates that there is a deficit of radio emission along
the length of the X-ray feature. These properties all argue that
feature~C corresponds a physical structure rather than just 
an enhancement in the brightness of the nebula. We propose
that feature~C corresponds
to a collimated outflow directed along the pulsar spin axis,
as is thought to be seen for the Crab and Vela pulsars and PSR~B0540--69
(\cite{hss+95}; \cite{gw00}; \cite{hgh01}; \cite{kma+01}),
and as is also suggested by the morphology of
the PWN in SNR~G0.9+0.1 (\cite{gpg01}).
The curvature of this outflow at large distances from the
pulsar is reflected in the overall curved morphology of the
surrounding nebula in this region (best seen in the 
{\em ROSAT}\ data of Fig.~\ref{fig_g320_most_rosat}).
This curvature, as also seen in the jet in the Crab Nebula,
presumably results from interaction with a gradient
in the ambient density and/or magnetic field. 

In the case of the Vela pulsar, Radhakrishnan \& Deshpande
(2001\nocite{rd01}) argue there is no true outflow along the
spin axis, and that the apparent ``jet'' feature is
a beaming effect produced by particles flowing along the
pulsar's magnetic axis.  However, in the case
of PSR~\psr\ the observed curvature of the jet and the
anti-correspondence with radio emission both 
provide good evidence for a physical cylindrical
structure which, to
persist over large spatial and temporal scales, can only be
directed along the spin axis.

As listed in Table~\ref{tab_spec}, feature~C has a photon index $\Gamma
\approx 1.7$, distinctly flatter than the value $\Gamma = 2.05\pm0.04$
determined for the diffuse surrounding component of the PWN.
This contrast in spectrum indicates that the conditions in 
feature~C must be distinct from those in the overall PWN:
either the site and mechanism for particle acceleration is
different in the outflow from that in the diffuse PWN, or
the synchrotron break in the outflow is at much higher energies.

We first consider the possibility that there is continuous
particle acceleration along the outflow, which generates
its harder spectrum. Such acceleration could be produced
by shocks and turbulence in the interface between the outflow
and its surroundings (\cite{hj88}; \cite{fbmo97}). 
However, if the jet is optically thin (which we
presume to be the case from its observed photon index)
then we expect it to be brightest along its edges,
not in its center as observed. We thus think this
possibility unlikely.

We therefore favor the alternative, that the
main source of injection and acceleration in
the outflow is the pulsar itself.
There is no good understanding of
how jets are produced in PWNe. For the purposes of
this discussion, we assume that
the injection spectra for the jet and for the overall
nebula are comparable. As discussed in \S\ref{sec_discuss_pwn},
multiple lines of evidence indicate that the overall PWN
cools from $\Gamma \sim 1.6$ to $\Gamma \approx 2.15$ due
to a synchrotron break at $\varepsilon_s \la 1$~keV. The
spectrum for the jet is comparable to the uncooled
spectrum for the diffuse PWN, indicating that any
synchrotron break in the jet is at an energy $\varepsilon_j> 10$~keV~$\gg
\varepsilon_s$.
We expect the magnetic field in the jet, $B_j$, to be significantly
higher than that in the surrounding PWN, $B_n \sim 8$~$\mu$G, in
order to support the jet against surrounding pressure.
Indeed we can estimate the minimum magnetic field in the outflow
by assuming equipartition. We assume the jet
to be a cylinder of length $l_j = 4' = 5.8$~pc and 
radius $r_j = 5'' = 0.12$~pc, corresponding to an emitting
volume $\sim8\times10^{54}$~cm$^3$. From the
spectral fit in Table~\ref{tab_spec}, we infer
an unabsorbed spectral luminosity at 5~keV 
$L_{5~{\rm keV}} = 1.3\times10^{33}$~erg~s$^{-1}$~keV$^{-1}$.
For a filling factor $\phi$ and a ratio of ion to
electron energy $\mu$,
equipartition arguments (\cite{pac70}, pp170--171) then imply a
minimum energy in the jet 
$E_j\sim4(1+\mu)^{4/7}\phi^{3/7}\times10^{44}$~erg and
a minimum magnetic field 
$B_j \sim 25(1+\mu)^{2/7} \phi^{-2/7}$~$\mu$G~$\gg B_n$.
The lifetime of synchrotron electrons at 5~keV is
then $t_j < 140(1+\mu)^{-3/7} \phi^{3/7}$~yr,
and so the rate at which the pulsar supplies
energy to the jet is $\dot{E}_j \sim E_j/t_j >
9(1+\mu)\times10^{34}$~erg~s$^{-1}$, corresponding
to $>0.5$\% of the pulsar's total spin-down luminosity.

The fact that the outflow shows an uncooled spectrum
in the \cxo\ band implies $\varepsilon_j \ga 10$~keV. This can only be true
if $t_f < 100(1+\mu)^{-3/7} \phi^{3/7}$~yr, where
$t_f$ is the flow time in the jet. We can thus
infer that the bulk velocity for the outflow
is $v_j = l_j/t_f \ga 0.2c$. The mean expansion
velocity for the overall PWN (assuming free
expansion into the SNR) is $\la0.05c \ll v_j$. This confirms
that the process by which the pulsar generates
the jet is energetically and dynamically 
distinct from the process by which the overall
pulsar wind is generated.

The high jet speed we have inferred here is in stark
contrast to the
situation in the Crab Nebula, whose polar outflow
has a velocity
$<0.03c$ (\cite{wag+01}).  While the nature of such jets
is uncertain, we note that in the model of Sulkanen \& Lovelace
(1990\nocite{sl90}), a strong jet forms in the pulsar magnetosphere
only when the neutron star's magnetic field is sufficiently strong.
We therefore hypothesize that the higher inferred dipole
magnetic field for PSR~\psr\ (a factor four higher than
for the Crab pulsar) may be an explanation for its 
much more energetic outflow.
The deficit of radio emission along the jet (seen in
Fig.~\ref{fig_g320_tongue}) is not
easily explained; the particle spectrum in the jet
may have a low-energy cutoff so that there are no
radio-emitting pairs (e.g.\ \cite{at93}).

The mean surface brightness\footnote{These values
are absorbed surface brightnesses, but can
be directly compared provided the absorption towards
both regions is comparable.}
for feature~C in the energy range 0.3--8.0~keV is
$\sim1\times10^{-7}$~photons~cm$^{-2}$~sec$^{-1}$.
On the other side of the pulsar, in a region to the NW of feature~E,
the mean brightness is
$\sim2\times10^{-8}$~photons~cm$^{-2}$~sec$^{-1}$.
Any counterpart to feature~C on the other
side of the pulsar is therefore $\ga5$ times fainter.

The absence of a counter-jet implies either that PSR~\psr\ is producing
an intrinsically one-sided outflow, or that a counter-jet is present
but cannot be seen.  While multipole components to the neutron star
magnetic field could indeed produce a one-sided jet (\cite{cbt96b}),
there is strong circumstantial evidence that the pulsar generates a
collimated outflow to its north through which  it is interacting with
the RCW~89 region (\cite{md83}; \cite{tkyb96}; \cite{bb97}; G99). This
conclusion is further bolstered by feature~D in
Figure~\ref{fig_g320_acisi}, which is significantly elongated along
the main axis of the system and widens with increasing distance from
the pulsar. We propose that this feature represents a jet which
produces no directly detectable emission, but for which we observe
enhanced emission in a cylindrical sheath along the interface between
the jet and its surroundings.
Assuming that feature~C and its unseen
counterpart have similar intrinsic surface brightnesses and outflow
velocities, relativistic Doppler boosting can account for the observed
brightness contrast provided that $v_j \cos \zeta \sim 0.28c$, where
$\zeta$ is the inclination of the outflow to the line-of-sight. Since
$\cos \zeta < 1$ for any inclination, we can infer a lower limit $v_j >
0.28c$, consistent with the estimate $v_j > 0.2c$ determined above
from the energetics of the system.

Brazier \& Becker (1997\nocite{bb97}) estimated $\zeta > 70^\circ$ on
the basis of {\em ROSAT}\ HRI data, in which the PWN appears to have a
``cross''-shaped morphology suggestive of an edge-on torus. However,
this inclination then results in an uncomfortably high velocity, $v_j >
0.8c$. In our \cxo\ observation, this cross-shaped morphology is not
apparent; it presumably resulted from imaging of the inner and
outer arcs (features E and 5 respectively) at the lower spatial
resolution and sensitivity of the {\em ROSAT}\ HRI.  A more reasonable
post-shock flow speed $v_j \approx c/3$ (e.g.\ \cite{kc84a})
corresponds to a smaller inclination angle $\zeta \sim 30^\circ$.
This low value of $\zeta$ has additional support
from radio polarization observations of PSR~\psr, which exclude the
larger values of $\zeta$ argued by Brazier \& Becker
(1997\nocite{bb97}) at the $\sim3\sigma$ level (\cite{cmk01}).

Models in which the pulsar's high-energy emission originates in outer
gaps of the magnetosphere result in a viewing angle $\zeta \sim
45^\circ$ (\cite{ry95}), while those in which the $\gamma$-ray emission
is produced in a polar cap generally require small values of $\zeta$
(\cite{khk+99}). We are thus unable to distinguish between these models
from the available data.  

\subsection{Interpretation of Features~E and 5}
\label{disc_bowshock}

Feature~E, the bright semi-circular arc to the north of the pulsar,
demarcates a clear transition between the collection of compact bright
features within $1'$ of the pulsar and the diffuse nebula which
extends to much larger scales. 
The arc-like morphology of this feature is suggestive of a bow-shock,
as would result where the ram-pressure from a fast-moving pulsar
balances the outflow from the relativistic pulsar wind. 
For a bow-shock to result, the pulsar's motion must be supersonic with
respect to the sound speed in the surrounding medium.
Since the PWN itself has a sound speed $c/\sqrt{3}$, this
condition can only be met if the reverse shock from the surrounding SNR has
collided with the PWN,
bringing thermal material to the center of the system (\cite{che98};
\cite{vagt01}; \cite{bcf01}). However, it is unlikely that
this stage in PWN evolution has yet occurred in this system,
since the resulting PWN would be brighter and more compact
than is observed, and
would occupy only a small fraction of the SNR's interior volume.

An alternative interpretation is suggested by the fact that
the dominant features seen in X-ray emission
from both the Crab and Vela PWNe 
are bright toroidal arcs surrounding the pulsar
(\cite{hss+95}; \cite{hgh01}). 
In these cases, it is thought that these tori
correspond to synchrotron-emitting
particles from a pulsar wind 
focused into the equatorial plane of the system.
Furthermore, in the case of the Crab, \cxo\ data show
a ring of emission situated interior to the torus (\cite{wht+00}),
which may represent the point where the free-flowing pulsar
wind first shocks.

In the case of PSR~\psr,
features~E and 5 both show an arc-like morphology which is
bisected by the symmetry axis of the nebula, similar to
what is seen in these other PWNe.
The orientation inferred from the
outflow being along the spin axis implies that features~E and 5 are closer
to the observer than is the pulsar,
and are produced by a wind whose line-of-sight velocity component is
directed towards us.
The upper limit on any departure from circularity in
the projected appearance of feature~E implies
$\zeta \la 30^\circ$, in agreement with the
estimate of the inclination angle determined 
in \S\ref{sec_discuss_orient}. Correcting for this inclination
angle, we can infer a separation 
$r_5 = 0.4-0.5$~pc between
the pulsar and feature~5,
and $r_E = 0.75-0.85$~pc
between the pulsar and feature~E.

\subsubsection{Analogy to the Crab Torus?}
\label{sec_disc_torus}

In interpreting this emission, we first note
that there are two characteristic time-scales associated
with such features: $t_{flow}$, the time taken for
particles to flow from the pulsar to this position,
and $t_{synch}$, the synchrotron lifetime for these
particles. 

We parametrize the upstream relativistic wind by the ratio, $\sigma_1$,
of the energy in electromagnetic fields to that carried in particles
(\cite{rg74}).  This implies that $\sigma_1 = B_1^2/4\pi\rho_1 \gamma_1
c^2$, where $B_1$ is the toroidal magnetic field, $\rho_1$ is the lab
frame total rest mass density, and $c\gamma_1 \gg c$ is the
four-velocity of the wind, all just upstream of the termination shock
in the pairs. We assume $\sigma_1 \ll 1$, and will show in
\S\ref{sec_wisps_struc} that this assumption is self-consistent.  In
this case, the bulk velocity in the post-shock flow at a distance $r$
from the pulsar is (\cite{kc84a}):
\begin{equation}
v(r) = \frac{c}{3} \left(\frac{r_s}{r}\right)^2,
\label{eq:vel}
\end{equation}
where $r_s$ is the distance from the pulsar to the termination
shock. We then find that: 
\begin{equation}
t_{flow} =  \frac{3 r_s}{c} \left( \frac{r}{r_s} \right)^3 =
4.9 \frac{r_s}{0.5 \; {\mathrm pc}} 
\left( \frac{r}{r_s} \right)^3 \; {\mathrm yr}.  
\label{eq:KC_flow_time} 
\end{equation}

For pairs emitting at energy $\varepsilon$~keV
in a magnetic field $B$~$\mu$G, the synchrotron
lifetime is
\begin{equation}
t_{synch} = 39~B^{-3/2}~\varepsilon^{-1/2}~{\rm kyr}.
\end{equation}

For the Crab Nebula, we adopt $r_s \approx 0.15$~pc and
$B \sim 100$~$\mu$G, and consider the Crab's X-ray
torus at $r \approx 0.4$~pc. At an energy $\varepsilon=5$~keV,
we find $t_{flow} \sim 30$~yr
and $t_{synch} \sim 20$~yr, so that these time scales
are comparable.
In the case of PSR~\psr, we assume that 
the location of the termination shock
corresponds approximately to that of feature~5 (the inner arc), 
and so adopt $r_s \approx r_5$. For
a magnetic field $B \sim B_n = 8$~$\mu$G,
we then find that in feature~E, $t_{flow} \sim 25$~yr~$\ll
t_{synch} \sim 800$~yr. Since the post-shock magnetic field
in the inner nebula is generally weaker than the mean
value for the PWN (\cite{kc84a}), this value of $t_{synch}$
is likely to be a lower limit, further widening the discrepancy
between the two time-scales.

Thus while the bright torus seen in the Crab simply corresponds to the
region in which most of the X-ray emitting particles radiate their
energy, synchrotron cooling is not
significant at this distance from PSR~\psr\ 
even at X-ray energies, and the brightness enhancement in
feature~E cannot be due to rapid dumping of pairs' energy into X-ray
photons. The long synchrotron lifetimes require that most of the X-ray
emission comes from a much larger volume, as is indeed observed.

It is therefore not valid to argue that feature~E is the analog of the X-ray
torus seen in the Crab; the much lower nebular magnetic
field in the case of PSR~\psr\ demands a different interpretation. 
This conclusion is supported by the fact that 
feature~E has a distinctly harder photon index than the
overall nebula and also shows a clear
radio counterpart (see Fig.~\ref{fig_g320_pol}), neither of
which would be expected if this feature were primarily
due to radiative losses.

\subsubsection{``Wisps'' in an Equatorial Flow}
\label{sec_disc_wisps}

The Crab Nebula
shows a series of toroidal optical and X-ray features close to the
pulsar, generally termed ``wisps'' (\cite{lam21};
\cite{sca69}; \cite{hss+95}; \cite{mhb+02}). These wisps form
at the termination shock, move outwards
in the equatorial plane at $\sim0.5c$, and then slow and fade
(\cite{ttt97}; \cite{hss+96}).
Furthermore, radio wisps have recently been
identified in the Crab Nebula, originating in the
same region as the wisps seen at higher energies,
and similarly moving outwards at high velocity (\cite{bfh01}). 
This result suggests that the 
radio, optical and X-ray emitting populations
are all accelerated in the same region, 
a result difficult
to explain in all models advanced to date for shock acceleration
at a PWN's wind termination shock.

As we have shown in \S\ref{sec_disc_torus},
the much lower magnetic field 
for the PWN around PSR~\psr\ results in $t_{synch} \gg t_{flow}$
for feature E, implying that it cannot
be a clone of the Crab's torus. 
In the following discussion, we therefore explore the consequences 
of identifying features~5 and E as analogs of the wisps seen 
in the equatorial outflow from the Crab.
The fact that we see in
Figure~\ref{fig_g320_pol} a clear radio counterpart to X-ray feature~E,
as is seen for the wisps in the Crab, 
gives phenomenological support to
its interpretation as a wisp-like feature,
while the high level of linear polarization is
consistent with the expected toroidal geometry for
the magnetic field in this region.
We note that because the synchrotron cooling time
is long compared to the flow time, thermal instabilities
due to synchrotron cooling, proposed by Hester (1998\nocite{hes98})
to account for the wisps in the Crab Nebula, cannot account
for features~E and 5 seen here. 

In the Crab Nebula, Gallant \& Arons (1994\nocite{ga94}, hereafter
GA94) extended the work of Kennel \& Coroniti (1984a\nocite{kc84a}) by
modeling the first optical wisp as emission from magnetized
electron/positron pairs, compressed by magnetically reflected heavy
ions just behind the leading edge termination shock in the pairs. These
ions form a highly accelerated Goldreich-Julian return current in the
flow.  For such magnetic reflection in the overall shock structure to
be a viable explanation of the wisps observed in the Crab, the ions
must have energy per particle comparable to the maximum energy
possible, $\gamma_i m_i c^2 \approx Z e \Phi_{open}$ (where $\gamma_i$,
$m_i$ and $Ze$ are the Lorentz factor, mass and charge of the ions
respectively, and $\Phi_{open}$ is the open field potential of the
pulsar's magnetosphere).  Subsequent time-dependent studies
(\cite{sa00}) have shown the time-variability due to the
electromagnetic instability of the reflected ion current to be
remarkably similar to the temporal behavior reported for the Crab's
inner X-ray ring (\cite{mhb+02}) and possibly for the first optical
wisp (\cite{ttt97}; \cite{hss+96}).

In the following sections, we propose that feature~E is an analog of
the Crab's second wisp, and results from compression
at the second, disorganized ion turning point in 
the pulsar's equatorial outflow.

\subsubsection{Structure of the Flow}
\label{sec_wisps_struc}

In the model of GA94, the upstream flow Lorentz
factor is:
\begin{equation}
\gamma_1 = \eta \frac{Ze\Phi_{open}}{m_i c^2} =
	   2.6 \times 10^6 \frac{\eta}{0.33} \frac{Z}{A} 
	   \left(\frac{\dot{E}}{1.8 \times 10^{37} \; {\rm erg~s}^{-1}}
	   \right)^{1/2},
\label{eq:gamma1}
\end{equation}
where $\eta$ is the fraction of the open field potential 
picked up by the ions, $A$ is the atomic number of the ions,
and the open magnetosphere's potential is
\begin{equation}
\Phi_{open} = \left( \frac{\dot{E}}{c} \right)^{1/2} 
	      = 7.4 \times 10^{15} \; {\rm V}.
\end{equation}
Previous uses of this model 
to study X-ray emission from the Crab,
PSR~B1957+20 and PSR~B1259--63 have consistently found
$\eta \sim 1/3$ to be the best choice
(\cite{at93}; GA94\nocite{ga94}; \cite{ta97}).
Soft X-ray observations suggest that hydrogen or possibly helium
dominate the composition of the upper layers of pulsar atmospheres
(\cite{pz00}), from which the ions of the return current are drawn.
We consequently adopt $Z\sim1$ and $A\sim1$ in further discussion.

The intrinsic thickness of the pair shock is comparable to
the Larmor radius in the pairs, $\mathcal{R}$$_{1\pm}$, based on the upstream 
Lorentz factor, $\gamma_1$, and upstream
magnetic field, $B_1=B_2/3$,
where $B_2$ is the magnetic field immediately downstream
of the pair shock (\cite{ghl+92}). Adopting $B_2 = 2.5$~$\mu$G 
(see Equation~[\ref{eq:post_shock_B}] below)
and $\gamma_1 \sim 2.6 \times 10^6$, we find that
$\mathcal{R}$$_{1\pm} \sim 5\times10^{-4}$~pc, corresponding
to an angular scale at 5~kpc of $0\farcs03$. The pair
shock creates a relativistic {\em thermal}\ distribution
of pairs (\cite{ghl+92}), with mean energy 
per particle $\sim \gamma_1 m_e c^2 \sim 1$~TeV,
whose synchrotron emission in the immediate post-shock
magnetic field is in the infrared, peaking around
5~$\mu$m and unobservably faint because of the slow cooling. The
observed non-thermal X-ray emission from feature~5 and at
larger radii is due
to the acceleration of pairs from the thermal pool. Therefore, in this
interpretation, feature~5 is the analog
of the first wisp in the Crab Nebula, and is created
by the deposition of the heavy ions' outflow momentum at the
first turning point in their magnetically reflected orbits.

Identifying feature~5 as an ion-driven compression implies
\begin{equation}
r_5 = \frac{1}{2} \mbox{$\mathcal{R}$}_{2i}
= \frac{1}{2} \frac{m_i c^2 \gamma_1}{ZeB_2} = 
\frac{1}{2} \frac{\eta \Phi_{open}}{B_2},
\end{equation}
where $\mathcal{R}$$_{2i}$ is the downstream ion Larmor radius.
Adopting $r_5 \approx 0.5$~pc, we then find that
\begin{equation}
3B_1 = B_2 = \frac{\eta \Phi_{open}}{2r_5} 
=2.5 \frac{\eta}{0.33} \; \mu{\mathrm G}.
\label{eq:post_shock_B}
\end{equation}

With a magnetic field at feature~5
of $\sim2.5$~$\mu$G, the Lorentz
factor of the X-ray emitting pairs is
$\gamma \approx 4\times10^8$, corresponding
to a Larmor radius $\mathcal{R}$$_{2\pm} \sim 0.09$~pc.
Therefore, the finite gyration of the pairs emitting
in X-rays will cause feature~5 to have an angular width
$\sim4''$ at an energy of 5~keV, no matter what the intrinsic
thickness of the structure in the magnetic field
and the pair plasma.  Figure~\ref{fig_g320_center} shows
that feature~5 is indeed $4''-5''$ across.
Most of the pairs in the flow have energy well below
the energies of the X-ray emitting particles.
In a distribution for which the number
of particles 
$N_\pm(\gamma) \propto \gamma^{-p}$, where $p=2\Gamma -1 \ga2$
these lower-energy particles are just more efficiently radiating
tracers with no substantial dynamical significance.

The radius of the next compression, $r_{next}$, can be
readily estimated. We expect $B(r_{next}) = k B(r_5) (r_{next}/r_5) $,
where the factor $ k>1$ represents the ion induced
compression of 
the magnetic field above the linear increase of $B$ with $r$; the linear
increase occurs 
simply because of deceleration of the expanding flow. GA94 found $k \sim 2$, 
while the time dependent simulations of Spitkovsky
\& Arons (2000\nocite{sa00}) show $k \sim 1.5$.

The radius of the second compression
occurs approximately at
\begin{equation}
r_{next} = r_5 + \mbox{$\mathcal{R}$}_{i}[B(r_{next})] = 
r_5 + \frac{2 r_5^2}{k r_{next}}.
\label{eq:spacing}
\end{equation}
Adopting $k \sim 1.5$ yields $r_{next}/r_5 = 1.76$, or
$r_{next} = 0.88~(r_5/0.5~{\rm pc})$~pc. The good
agreement between $r_{next}$ and
the observed radius for feature~E, $r_E = 0.75-0.85$~pc,
supports identifying feature~E as the second ion
compression in the flow. Relativistic cyclotron instability
of the reflected ions (\cite{sa00} and in preparation)
disorganizes their flow at larger radii, and no further coherent
compressions (a.k.a ``wisps'') at still larger radii
can be produced by this model, in agreement with the observations.

The model predicts the peak magnetic field in feature~E to be
\begin{equation}
B_E = kB_2 \frac{r_E}{r_5} 
\sim 6.5 \frac{0.5 {\mathrm pc}}{r_5} \frac{\eta}{0.33} \; \mu{\mathrm G}; 
\label{eq:peak_B_E}
\end{equation}
the average magnetic field in feature~E is closer to 5~$\mu$G.

\subsubsection{Energetics of the Flow}
\label{sec_disc_energy}

The energetics of the ion flow can be estimated from theory,
since the electrodynamics of the magnetosphere constrain their
outflow rate to the Goldreich-Julian value,
\begin{equation}
\dot{N}_i = \frac{2 \Phi_{open}c}{Ze} = 3 \times 10^{33} \;
{\mathrm s^{-1}}.
\label{eq:GJ_loss}
\end{equation}
From Equation~(\ref{eq:GJ_loss}), the ion
density just upstream of the shock in pairs\footnote{Assuming $r_s$
is not much less than the radius of the first turning 
point in the ions' orbits, as is true for the GA94 model of the Crab.}
is then
\begin{equation}
n_{1i} = \frac{\dot{N}_i}{4\pi r_5^2 c} \frac{4\pi}{\Delta \Omega}
       = 3.3 \times 10^{-15} \frac{4\pi}{\Delta \Omega} \; {\mathrm cm}^{-3},
       \end{equation}
where $\Delta \Omega$ is the solid angle subtended by the ion
outflow. In the Crab, the appearance of the wisps and of
the X-ray torus suggest $4\pi / \Delta \Omega \sim 5.8$. We
do not know whether the equatorial outflow from PSR~\psr\ is
similarly flattened, although the presence
of the jet (feature~C) clearly indicates that
some degree of flattening is present.

The most meaningful way to characterize the ions' energetics is to
compare them with the electromagnetic energy flux just upstream
of the pair shock, by computing the parameter $\sigma$ for the ions.
We find
\begin{equation}
\sigma_{1i}^{\rm (theoretical)} = \frac{B_1^2}{4\pi n_{1i} \gamma_1 m_i c^2}  
 = 8\times 10^{-4}~\frac{0.33}{\eta}~\frac{5.8}{\frac{4\pi}{\Delta 
 \Omega}}~\left(\frac{r_5}{0.5~{\rm pc}}\right)^2.
 \label{eq:sigma_i}
\end{equation}

Theory suggests that PSR~\psr\ should produce $\kappa_\pm \approx 10^3$
electron/positron pairs per primary particle in the polar
current-carrying beam within the magnetosphere 
if the surface magnetic field is strictly dipolar (\cite{ha01});
$\kappa_\pm$ might be a few times larger if the surface
magnetic field is strongly distorted (\cite{ba82}; \cite{ha01})
or if the dipole is strongly offset (\cite{aro98}). 
Then
\begin{equation}
\dot{N}_\pm^{\rm (theoretical)} = \kappa_\pm \dot{N}_i 
   = 3 \times 10^{36} \frac{\kappa_\pm}{10^3} \; {\mathrm s}^{-1}.
   \label{eq:pair_rate_theoretical}
\end{equation}

The ion compressions that lead to the observed
wisps only occupy an equatorial sector of latitudinal extent
$\Delta\Omega$ where the electric return current from the pulsar flows, 
while the pairs from the star flow out along field lines which
fill the whole space around the star, in the polar as well as the 
equatorial sectors.  Therefore, the pairs expand quasi-spherically so that
\begin{equation}
n_{1\pm} = n_{1+} + n_{1-} = 2 \kappa_\pm n_{1i} \frac{\Delta \Omega}{4\pi}
     = 6.9 \times 10^{-12}~\frac{\kappa_\pm}{10^3} \; {\mathrm cm}^{-3} ,
\end{equation}
and so
\begin{equation}
\sigma_{1\pm}^{\rm (theoretical)} = \frac{B_1^2}{4\pi n_{1\pm} \gamma_1 m_e c^2} 
= 4\times 10^{-3}~\frac{0.33}{\eta}~\left(\frac{r_5}{0.5~{\rm pc}}\right)^2;
\label{eq:pair_sigma_theory} 
\end{equation} 
the ions have five times the flow energy of the pairs.

Since
$$
\frac{1}{\sigma_1} = \frac{1}{\sigma_{1\pm}} + \frac{1}{\sigma_{1i}},
$$
we find that
\begin{equation}
\sigma_1^{(\mathrm theoretical)} = 
7\times10^{-4}~\frac{\eta}{0.33}~\frac{r_5}{0.5~{\rm pc}}, 
\label{eq:sigma_total} 
\end{equation}
where we have assumed $\kappa_\pm = 10^3$.

Estimating the pair density and thus $\sigma_{1\pm}$ from
X-ray data is more difficult, since the
relatively narrow bandwidth of the observations means that
we have very little constraint on the overall
energetics. The spectral fits to feature~E
listed in Table~\ref{tab_spec} indicate
an unabsorbed 0.5--10~keV flux density
$f_E \approx 3.6 \times 10^{-12}$~erg~cm$^{-2}$~s$^{-1}$
and a photon index $\Gamma = 1.7$. For a distance
to the source of 5~kpc, these imply
a spectral luminosity in the \cxo\ band of 
\begin{equation}
L_\varepsilon = 2.8 \times 10^{33}~\varepsilon^{-0.7}~{\rm 
erg~s^{-1}~keV^{-1}}.
\label{eq:obs_flux}
\end{equation}

The power law form of the spectrum of feature~E suggests synchrotron
radiation from a power law distribution of particles. From synchrotron
theory,
the isotropic emission rate  of an isotropic power law distribution of
electrons and positrons with energy spectrum
$N_\pm (E) = K E^{-p}, \; E_{\rm min} < E < E_{\rm max}$ is
(\cite{hje99})
\begin{equation}
q_\varepsilon = 1.1 \times 10^{-14}~K B_E^{1.7} \varepsilon^{-0.7}~{\rm 
erg~cm^{-3}~s^{-1}~keV^{-1}},
\label{eq:synch_flux}
\end{equation}
where we have used $p = 2\Gamma - 1 = 2.4$.

We model feature E as occupying a volume
$V_E = 4\pi r_E^2 \Delta r (\Delta \Omega /4\pi)$,  where
$\Delta r$ is the radial width of feature~E. Adopting
$r_E = 0.8$~pc, $\Delta r \approx 0.4$~pc and
$4\pi/\Delta\Omega = 5.8$, we find $V_E = 1.6\times10^{55}$~cm$^3$.
Since $L_\varepsilon = V_E \, q_\varepsilon$, we can
combine Equations~(\ref{eq:obs_flux}) and (\ref{eq:synch_flux}) 
to find that
$K = 1.0\times10^{-9}~{\rm erg}^{1.4}~{\rm cm}^{-3}$,
where we have used $B_E=5$~$\mu$G.
Integrating the power law particle 
spectrum over energy yields a total pair density in feature E of
\begin{equation}
n_\pm^{(E)} = n_+^{(E)} + n_-^{(E)} =  2.3 \times 10^7
E_{\rm min}^{-1.4}~{\rm cm^{-3}},
\end{equation}
where $E_{\rm min}$~eV is the minimum energy of the
power-law particle distribution.
The injection rate into feature~E is therefore
\begin{equation}
\dot{N}_\pm^{(E)} = 4\pi r^2 \frac{\Delta \Omega}{4 \pi} n_\pm^{(E)} v(r_E)
   = 1.2 \times 10^{54}~E_{\rm min}^{-1.4}~{\rm s}^{-1}.
\label{eq:E_injection}
\end{equation}
As per Equation~(\ref{eq:vel}),
we used $v(r_E) = (c/3) (r_5 /r_E)^2 = 3.9 \times 10^9$~cm~s$^{-1}$
in the numerical evaluation.

We have no observational handle on $E_{\rm min}$, other than it must be small
enough to allow the synchrotron critical energy 
$(3/2) (\hbar eB /m_e c) (E_{\rm min} /m_e c^2 )^2 $ to be well below the
lower limit of the effective Chandra band for this source, 
$\varepsilon \sim 1$~keV. Therefore, observationally 
$E_{\rm min} < 53$~TeV.
Theoretically,
existing work on the particle acceleration properties of
quasi-transverse relativistic magnetosonic shocks suggests the downstream
power law should flatten (and perhaps cut off) below the energy
$\gamma_1 m_e c^2$ (\cite{kc84}; \cite{hagl92}).
From Equation~(\ref{eq:gamma1}), we then would have 
$E_{\rm min} \geq 1.3 (\eta/0.33)$~TeV.  The particle
acceleration theory suggests the true power form of the particle 
distribution sets in at two to three times this energy, in which case 
\begin{equation}
\dot{N}_\pm^{(E)} \sim 4 \times 10^{36} \left(
\frac{\eta}{0.33}\right)^{-1.4}~{\rm s}^{-1}
\label{eq:pair_rate_observational} 
\end{equation}
which compares well with Equation~(\ref{eq:pair_rate_theoretical}). 

Equation~(\ref{eq:pair_rate_observational}) yields:
\begin{eqnarray}
\sigma_{1\pm}^{(E)} = \frac{B_1^2}{4\pi n_{1\pm} 
\gamma_1 m_e c^2} =  
    \frac{B_1^2 r_5^2 c}{\dot{N}_\pm \gamma_1 m_e c^2} \nonumber \\
	 \approx 7\times10^{-3} \left(\frac{\eta}{0.33} \right)^{2.4},
\end{eqnarray}
a similar value to the theoretical estimate of 
Equation~(\ref{eq:pair_sigma_theory}).
From either approach $\sigma_{1\pm}$ is small,
and is comparable to the value $\sigma \approx 3\times10^{-3}$
inferred for the Crab Nebula (\cite{kc84a}; \cite{ec87}).

Feature~E forms less than half a complete torus ---
this arc is 3--5 times brighter than any counterpart
on the other side of the pulsar. 
The flow velocity given by Equation~(\ref{eq:vel}) is
too small at feature~E to invoke Doppler boosting as 
an explanation (cf.\ \cite{ppp+87}).  
We thus can offer no
explanation for this asymmetry, except to note that
a similar situation is seen in the equatorial structures
in the Vela~PWN and in SNR~G0.9+0.1 (\cite{hgh01}; \cite{gpg01}).

Nevertheless, we believe that
the ideas developed by GA94\nocite{ga94} and Spitkovsky \& Arons
(2000\nocite{sa00}) to explain the spatial (and temporal)
structure of the equatorial interaction zone between
the Crab Nebula and its pulsar can be usefully applied
to the equatorial interaction
between PSR~\psr\ and its PWN, as revealed by the
\cxo\ observations presented in this paper. To the
extent that the model outlined here does represent the facts,
the basic conclusion is the same as was the case
in the GA94 model for the Crab, namely $\sigma_1 \ll 1$. This
result has been extracted from a close study of the
detailed structure of the interaction zone rather than
from modeling the global dynamics of the whole PWN
(cf.\ \cite{kc84a}), the latter of which is difficult
to apply to the complicated morphology surrounding PSR~\psr.

We note that if features~E and 5 are indeed analogs
of the wisps seen in the Crab, then we similarly expect
them to move away from the pulsar at high velocity.
For example, motion outwards at $0.5c$ would
correspond to proper motions of a few arcsec per year at
the distance of PSR~\psr, which could easily be detected by \cxo\
at subsequent epochs.

\subsection{Small Scale Structure Near the Pulsar}

The high resolution of \cxo\ has revealed emission from several small-scale
features close to the pulsar, as seen in
Figure~\ref{fig_g320_center}.

Feature~6 is likely to correspond to emission from the O~star
Muzzio~10 (\cite{muz79}; \cite{om80}).
X-ray emission from an O6.5III star can be typically
approximated by a Raymond-Smith spectrum with $kT\approx 0.2-0.5$~keV
and an unabsorbed luminosity (0.3--10.0~keV) of $\sim
1-3\times10^{32}$~erg~s$^{-1}$ (\cite{bsc96}), corresponding
to an unabsorbed flux density (0.3--10.0~keV) $f_x \sim
(0.4-4)\times10^{-13}$~erg~cm$^{-2}$~s$^{-1}$ for a photometric
distance in the range 2.5--4.6~kpc (\cite{shmc83}; \cite{are91}).
The crude spectral
parameters inferred for this source in Table~\ref{tab_spec} are
consistent with these values.

What features 1--4 represent is not immediately clear. If we accept
the argument made in \S\ref{sec_disc_torus} that $r_s \approx r_5$, 
then features 1--3 (and possibly feature~4, depending
on projection effects) originate in the zone in which the wind is
still freely expanding.

In the Crab Nebula, a variety of small-scale
structures has similarly been identified
within the unshocked wind zone.
``Knot 1'' and ``knot 2'' (the latter of which is
also termed ``the sprite'') are resolved optical structures
lying 1500 and 9000~AU, respectively, from the Crab pulsar
along the jet axis (\cite{hss+95}).
It has been proposed that these features
correspond to quasi-stationary shocks
in the polar outflow from the pulsar (\cite{msh+96}; \cite{lou98}),
or, in the case of knot~1, to a
sheath of emission surrounding this outflow (\cite{gsh+96}).
The two knots both show significant variability in their
brightness, position and morphology on time scales of days
to months (\cite{hss+96}; \cite{hes98}).
Similarly time-variable 
knots of emission are also seen in X-rays close
to the Vela pulsar (\cite{pksg01}).

In the following discussion, we focus on feature~1,
the knot of emission closest to PSR~\psr. We estimate 
feature~1 to have approximate dimensions $3''\times5''$,
and its projected separation from the pulsar to be 
$r_{\rm knot} \approx 4'' =
0.1$~pc. We assume that this emission is generated
by particles accelerated in some localized turbulent region,
such as might be produced by the collision of inhomogeneous
wind streams proposed by Lou (1998\nocite{lou98}). The
spectrum observed for feature~1 ($\Gamma \sim 1.2$) is
somewhat harder than the uncooled spectrum for the overall PWN
($\Gamma \sim 1.6$), supporting the possibility 
that this is a region of separate particle acceleration.
In this case,  we
can interpret the extent of this feature as corresponding
to the relativistic Larmor radius of gyrating pairs. Although
the number of photons available is limited, there is the suggestion
in the data that the extent of the knot increases slightly with increasing
photon energy, consistent with this interpretation.  At a photon
energy $\varepsilon$~keV and in a magnetic field $B_{\rm knot}$~$\mu$G, the
pair Larmor radius is $\mathcal{R}$$_{\rm knot} = 
0.13(\varepsilon/B_{\rm knot}^3)^{1/2}$~pc. 
Adopting $\mathcal{R}$$_{\rm knot} \sim 0.05$~pc and 
$\varepsilon=5$~keV, we can infer $B_{\rm knot} \sim 3$~$\mu$G.

We computed in \S\ref{sec_disc_energy} that the rate
of pair production in the pulsar wind is $\dot{N}_\pm \approx
4\times10^{36}$~s$^{-1}$. The pair density in feature~1
is therefore
\begin{equation}
n_{\pm, \rm knot} = \frac{\dot{N}_\pm}{4\pi r_{\rm knot}^2 c} 
		   \approx 1.1 \times 10^{-10}~{\rm cm}^{-3}.
\end{equation}
In Equation~(\ref{eq:gamma1}) we computed an upstream flow
Lorentz factor $\gamma_1 = 2.6 \times10^6$. Combining
these estimates for $B_{\rm knot}$, $n_{\pm, \rm knot}$
and $\gamma_1$, we can infer by analogy with
Equation~(\ref{eq:pair_sigma_theory}) that
$\sigma_\pm < 3 \times 10^{-3}$
in this region. (We adopt this value as an upper limit, since
compression and turbulence likely enhance the magnetic
field in this feature above the ambient value.)

This low value of $\sigma$ is consistent with the various models
for energy transport in the unshocked wind, which generally
require the transition from $\sigma \gg 1$ (at the light cylinder)
to $\sigma \ll 1$ (at the termination shock) to occur  at $r \ll
r_{\rm knot}$ (\cite{cor90}; \cite{mic94}; Melatos, private communication).
It is interesting to note that in the large-amplitude plasma wave model
proposed by Melatos (1998\nocite{mel98}), the wind beyond
this transition point evolves as $\sigma \propto r^2$. We thus
expect $\sigma$ for feature~1 to be $\sim$20 times smaller
than that inferred at the termination shock, a result not inconsistent
with our calculations.

A better insight into the nature of these compact features will require
observations of this system at multiple epochs and at longer
wavelengths (e.g.\ in the near-infrared).  
With such data, we will be better able to determine the
broadband spectrum and hence total energetics of these sources, can
establish whether these features are persistent or transient, and can
determine whether any of these sources shows outward proper motion which
could associate them directly with outflow from the pulsar.  Indeed,
\cxo\ has identified significant changes in the positions and
brightnesses of small-scale structure in the Vela PWN on a time-scale
of seven months (\cite{pksg01}).

\section{Summary and Conclusions}

Our \cxo\ observations have confirmed that PSR~\psr\ interacts
with its surroundings in a spectacular and highly anisotropic fashion. 
Our main results are as follows:

\begin{enumerate}

\item PSR~\psr\ powers a large elongated non-thermal nebula,
whose main symmetry axis most likely corresponds to the pulsar spin axis.
A variety of arguments implies a mean nebular magnetic field
of $B_n\sim8$~$\mu$G, corresponding to a synchrotron break
just below the X-ray band. The high sensitivity of these
data has allowed us to map out the full extent of the X-ray
PWN. We have shown that this nebula matches well the extent
of a previously-identified ``tongue'' of radio emission near the pulsar,
and propose this latter feature to be the radio nebula powered by
pulsar.

\item We confirm previous claims of a one-sided collimated outflow 
emanating from the pulsar along the main symmetry axis,
and which generates a channel of reduced radio emission.
We determine that this jet carries away
at least 0.5\% of the pulsar's spin-down luminosity, and
that the bulk velocity in the outflow is $v_j>0.2c$.
We have identified an elongated sheath on the other side
of the pulsar which indicates the existence of a counter-jet,
despite the lack of direct emission from such a source.
The brightness contrast between the two outflows can be explained
by Doppler boosting, and implies an inclination between
the pulsar spin axis and the line-of-sight $\zeta\sim30^\circ$.

\item Approximately 0.5 and 0.8~pc to the north of the pulsar,
we identify two arcs of emission with a toroidal geometry
which similarly implies an inclination $\zeta\sim30^\circ$.
A re-analysis of radio observations of the region shows that
the second of these arcs has a clear radio counterpart and
is strongly linearly polarized. We show that the flow
time in these features is much less than the time scale
for radiative losses, and that they therefore cannot
correspond to a radiating torus as seen in the Crab Nebula.
Rather, we interpret these features as ``wisps'' in
an equatorial flow, and show
that such an assumption is consistent with these
sources representing internal structure in the termination
shock of a particle-dominated
ion-loaded wind, as proposed by Gallant \& Arons
(1994\nocite{ga94}) to explain the wisps in the Crab Nebula. 
We infer a ratio at the shock of Poynting to
particle flux for electron/positron pairs $\sigma_\pm \sim 0.005$,
a value similar to that argued for the Crab.

\item We identify several compact knots of emission less
than 0.5~pc from the pulsar, within the unshocked wind zone.
The hard spectrum seen for these knots suggests a different
acceleration mechanism than for larger scale nebular features beyond
the termination shock. We infer a ratio of Poynting to particle
flux for one of these knots $\sigma_\pm < 0.003$, indicating
that the transition from a magnetically-dominated to a particle-dominated
wind occurs less than 0.1~pc from the pulsar.

\end{enumerate}

As has been the case in the past, new data on this source
raise as many
questions as they answer. We lack an understanding of
how PSR~\psr\ generates such a fast-moving and collimated
outflow over such large scales, nor is it clear why
such a striking feature is not seen in most other PWNe.
We have also not addressed the nature of the interaction
between this outflow and the surrounding SNR; a forthcoming paper
will present an analysis of X-ray emission from the \rcw\ region,
which may better elucidate the physical conditions associated
with this process. While we have outlined a self-consistent
model in which the arcs seen near the pulsar are wisps as
in the Crab, we have no explanation for
why these features are only seen on one side of the
pulsar (it is important to realize that the
nature of the Crab's wisps are themselves still a matter of heated debate).
We clearly need to extend our coverage to other wavelengths and
additional epochs if we are to better understand the observed structures.
Finally, the process which produces the innermost X-ray features
is completely unknown, and will also require further study.

To conclude, we comment that for many decades the Crab
Nebula has been the focus of almost
all efforts to understand the process by which pulsars couple
to their environment. With data such as
those presented here, we can now finally 
extend such studies to a wider sample of sources.

\begin{acknowledgements}

We thank Fred Baganoff, Jonathan McDowell, Paul Plucinsky, Mallory
Roberts and Randall Smith for advice on data analysis,  Matthew Baring,
Jeff Hester, Maxim Lyutikov and Andrew Melatos for useful discussions,
and the anonymous referee for a careful reading of the manuscript. B.
M. G. also thanks the Australia Telescope National Facility for their
hospitality while some of this paper was written.  This work was
supported by NASA through SAO grant GO0-1134X, Hubble Fellowship grant
HST-HF-01107.01-A (B.M.G.), LTSA grant NAG5-8063 (V.M.K.) and contracts
NAS8-37716 and NAS8-38252 (M.J.P.).  B.M.G. also acknowledges the
support of a Clay Fellowship awarded by the Harvard-Smithsonian Center
for Astrophysics.

\end{acknowledgements}


\clearpage

\begin{table}
\caption{Detected count rates for selected regions.
Uncertainties are  $\pm1\sigma$.\label{tab_rates}}
\begin{tabular}{lcc} \hline \hline
Region & Count Rate\tablenotemark{a} & 
Mean Surface Brightness\tablenotemark{a} \\ 
& (cts~s$^{-1}$) & ($10^{-5}$ 
cts~s$^{-1}$~arcsec$^{-2}$) \\ \hline
Entire source & $8.79\pm0.03$ & $2.04\pm0.01$  \\
PSR~\psr\ & $0.140\pm0.003$\tablenotemark{b} & $\ldots$ \\
\rcw & $2.39\pm0.01$ & $3.51\pm0.01$  \\
C  & $0.212\pm0.005$ & $3.38\pm0.08$  \\
E  & $0.146\pm0.005$ & $3.6\pm0.1$  \\
1 & $0.014\pm0.001$ & $78\pm6$  \\
2 & $0.007\pm0.001$ & $29\pm4$  \\
3 & $0.042\pm0.002$ & $86\pm4$  \\
4 & $0.029\pm0.002$ & $66\pm5$  \\
5 & $0.068\pm0.003$ & $15\pm1$  \\ 
6 & $0.0017\pm0.0004$ & $\ldots$ \\ \hline \hline
\end{tabular}

\tablenotetext{a}{Count rates and surface brightnesses are those
detected over the energy range 0.5--10~keV,
and have been corrected for background. Surface brightnesses
are only given for extended sources.}
\tablenotetext{b}{The pulsar suffers from severe pile-up;
this value is therefore a significant underestimate of
the incident count rate.}
\end{table}

\begin{table}
\caption{Spectral fits to various sub-regions of the source.\label{tab_spec} }
\begin{tabular}{lccccccccc} \hline \hline
Region & Model & $N_H$ ($10^{21}$~cm$^{-2}$) &
$\Gamma$ / $kT$ (keV) &
$f_x$ ($10^{-12}$~erg~cm$^{-2}$~s$^{-1}$)\tablenotemark{a} &
$\chi_\nu^2/\nu$ \\ \hline
Diffuse PWN
  & PL & $9.5\pm0.3$ & $2.05\pm0.04$ & $55\pm3$ & $1686/1268=1.33$ \\ \hline
C & PL & $8\pm1$ & $1.6\pm0.1$ &
$5\pm1$ & $212/237=0.89$ \\
 & PL & 9.5 (fixed) & $1.70\pm0.09$ & 
 $5.2\pm0.5$ & $217/238=0.91$ \\ \hline
E & PL & $8^{+2}_{-1}$ & $1.6\pm0.2$ &
 $3\pm1$ & $206/193=1.07$ \\
 & PL & 9.5 (fixed) & $1.7\pm0.1$ &
 $3.6\pm0.4$ & $207/194=1.07$ \\ \hline
1 & PL & $10^{+8}_{-7}$ & $1.3^{+0.7}_{-0.5}$ &
 $0.4^{+0.7}_{-0.4}$ & $24/21=1.13$ \\
 & PL & 9.5 (fixed) & $1.2\pm0.3$ &
 $0.4\pm0.1$ & $24/22=1.08$ \\ \hline
2 & PL & $2^{+20}_{-2}$ & $0.6^{+1.2}_{-0.6}$  &
 $\sim0.2$ & $12/11=1.10$ \\
 & PL & 9.5 (fixed) &  $1.0^{+0.6}_{-0.5}$ & 
 $\sim0.2$ & $13/12=1.06$ \\ \hline
3 & PL & $11^{+3}_{-2}$ & $1.7\pm0.3$ &
 $1.0^{+0.6}_{-0.3}$ & $57/61 = 0.93$ \\
 & PL & 9.5 (fixed) & $1.6\pm0.1$ & 
 $1.0\pm0.2$ & $57/62 = 0.92$ \\ \hline
4 & PL & $11^{+5}_{-4}$ & $1.3\pm0.4$ &
 $0.8^{+1.4}_{-0.8}$ & $50/52 = 0.96$ \\
 & PL & 9.5 (fixed) & $1.3\pm0.2$ &
 $0.8\pm0.2$ & $50/53 = 0.95$ \\ \hline
5 & PL & $6\pm2$ & $1.3^{+0.3}_{-0.2}$ &
 $1.6^{+0.6}_{-0.5}$ & $101/84 = 1.21$ \\
 & PL & 9.5 (fixed) & $1.6\pm0.1$ & 
 $1.6\pm0.2$ & $105/85 = 1.24$ \\ \hline
6 & RS & $\sim$10 &  $\sim0.6$ & $\sim$0.1 & $\ldots$
\\ \hline \hline
\end{tabular}

\tablenotetext{}{Uncertainties are all at 90\% confidence.
All models assume interstellar absorption using the
cross-sections of Ba\protect\mbox{\l}uci\protect\'{n}ska-Church 
\& McCammon (1992\protect\nocite{bm92}), assuming solar abundances.
Models used: ``PL'' indicates a power law of the
form $f_\varepsilon \propto \varepsilon^{-\Gamma}$ where $\Gamma$ is the
photon index;
``RS'' indicates a Raymond-Smith spectrum of
temperature $T$ (\protect\cite{rs77}).}
\tablenotetext{a}{Flux densities are for the energy range 0.5--10~keV,
and have been corrected for interstellar absorption.}
\end{table}

\clearpage

\end{document}